\documentclass[twocolumn]{autart}    

\usepackage{graphicx}          

\usepackage{amssymb,amsmath,amsfonts}
\usepackage{bm}
\usepackage{color}
\usepackage{cite}
\usepackage{booktabs}
\usepackage{threeparttable}
\usepackage{enumerate}
\usepackage{algorithmicx}
\usepackage{algorithm}
\usepackage{algpseudocode}
\usepackage{nomencl}
\usepackage{commath}
\usepackage{makecell}
\usepackage{blindtext}
\usepackage{soul}
\usepackage{comment}
\usepackage{bm}
\usepackage{mathtools}
\usepackage[svgnames]{xcolor}
\usepackage{tikz}
\usepackage{derivative}
\usepackage[export]{adjustbox}
\usepackage[siunitx, RPvoltages]{circuitikz}
\usetikzlibrary{positioning, shapes}
\usepackage{url}
\usepackage{accents}
\usepackage[super]{nth}
\usepackage{mathrsfs}
\usepackage{verbatim}
\usepackage{enumitem}
\usepackage{tgcursor}  
\usepackage[T1]{fontenc}
\usepackage{stackengine}
\usepackage{nomencl}
\makenomenclature
\DeclareMathAlphabet\euscr{U}{eus}{m}{n}
\DeclareMathAlphabet\urwscr{U}{urwchancal}{m}{n}

\def\XXint#1#2#3{{\setbox0=\hbox{$#1{#2#3}{\int}$}
     \vcenter{\hbox{$#2#3$}}\kern-.5\wd0}}

\newcommand{\vect}[1]{\mathbf{#1}}
\newcommand{\matr}[1]{\mathbf{#1}}
\newcommand{\tran}{\mathsf{T}}

\newcommand{\unaryminus}{-}

\DeclareMathAlphabet{\mathbbmsl}{U}{bbm}{m}{sl}
\newsavebox{\foobox}

\newcommand\eqm[1]{\overline{#1}}

\DeclareMathAlphabet\EuRoman{U}{eur}{m}{n}
\SetMathAlphabet\EuRoman{bold}{U}{eur}{b}{n}

\makeatletter
\renewcommand*\env@matrix[1][*\c@MaxMatrixCols c]{%
  \hskip -\arraycolsep
  \let\@ifnextchar\new@ifnextchar
  \array{#1}}
\makeatother

\def\XXint#1#2#3{{\setbox0=\hbox{$#1{#2#3}{\int}$ }
\vcenter{\hbox{$#2#3$ }}\kern-.6\wd0}}

\makeatletter
\renewcommand*\env@matrix[1][\arraystretch]{%
  \edef\arraystretch{#1}%
  \hskip -\arraycolsep
  \let\@ifnextchar\new@ifnextchar
  \array{*\c@MaxMatrixCols c}}
\makeatother

\makeatletter
\DeclareFontFamily{OMX}{MnSymbolE}{}
\DeclareSymbolFont{MnLargeSymbols}{OMX}{MnSymbolE}{m}{n}
\SetSymbolFont{MnLargeSymbols}{bold}{OMX}{MnSymbolE}{b}{n}
\DeclareFontShape{OMX}{MnSymbolE}{m}{n}{
    <-6>  MnSymbolE5
   <6-7>  MnSymbolE6
   <7-8>  MnSymbolE7
   <8-9>  MnSymbolE8
   <9-10> MnSymbolE9
  <10-12> MnSymbolE10
  <12->   MnSymbolE12
}{}
\DeclareFontShape{OMX}{MnSymbolE}{b}{n}{
    <-6>  MnSymbolE-Bold5
   <6-7>  MnSymbolE-Bold6
   <7-8>  MnSymbolE-Bold7
   <8-9>  MnSymbolE-Bold8
   <9-10> MnSymbolE-Bold9
  <10-12> MnSymbolE-Bold10
  <12->   MnSymbolE-Bold12
}{}

\let\llangle\@undefined
\let\rrangle\@undefined
\DeclareMathDelimiter{\llangle}{\mathopen}%
                     {MnLargeSymbols}{'164}{MnLargeSymbols}{'164}
\DeclareMathDelimiter{\rrangle}{\mathclose}%
                     {MnLargeSymbols}{'171}{MnLargeSymbols}{'171}
\makeatother

\DeclareFontFamily{U}{matha}{\hyphenchar\font45}
\DeclareFontShape{U}{matha}{m}{n}{
      <5> <6> <7> <8> <9> <10> gen * matha
      <10.95> matha10 <12> <14.4> <17.28> <20.74> <24.88> matha12
      }{}
\DeclareSymbolFont{matha}{U}{matha}{m}{n}
\DeclareFontSubstitution{U}{matha}{m}{n}

\DeclareFontFamily{U}{mathx}{\hyphenchar\font45}
\DeclareFontShape{U}{mathx}{m}{n}{
      <5> <6> <7> <8> <9> <10>
      <10.95> <12> <14.4> <17.28> <20.74> <24.88>
      mathx10
      }{}
\DeclareSymbolFont{mathx}{U}{mathx}{m}{n}
\DeclareFontSubstitution{U}{mathx}{m}{n}

\DeclareMathDelimiter{\vvvert}{0}{matha}{"7E}{mathx}{"17}

\definecolor{myblue}{RGB}{0, 0, 0}

\begin{document}

\begin{frontmatter}

\title{
A reference frame--based microgrid primary control for ensuring global convergence to a periodic orbit} 

\thanks[footnoteinfo]{
Corresponding author Y.~Li. Tel. +1-814-863-9571. 
Fax +1-814-865-6392. This work was supported by Office of Naval Research under the award N00014-22-1-2504.}

\author[one]{Xinyuan Jiang}\ead{xuj49@psu.edu},    
\author[one]{Constantino M. Lagoa}\ead{cml18@psu.edu},               
\author[two]{Daning Huang}\ead{daning@psu.edu},  
\author[one]{Yan Li\thanksref{footnoteinfo}}\ead{yql5925@psu.edu}  

\address[one]{Department of Electrical Engineering, Pennsylvania State University, University Park, PA 16802, USA}  
\address[two]{Department of Aerospace Engineering, Pennsylvania State University, University Park, PA 16802, USA}             


\begin{abstract}                          
Power systems with a high penetration of renewable generation are vulnerable to frequency oscillation and voltage instability. Traditionally, the stability of power systems is considered either in terms of local stability or as an angle oscillator synchronization problem with the simplifying assumption that the dynamics of the amplitudes are on much shorter time scales. Without this assumption, however, the steady state being studied is essentially a limit cycle with the convergence of its orbit in question. In this paper, we present a method to analyze the orbital stability of a microgrid and propose a voltage controller for the inverter-interfaced renewable generators. The main hurdle to the problem lies in the constant terms in the rotating internal reference frames of each generator. We extend the shifted passivity of port-Hamiltonian systems to the analysis of limit cycles and prove that, if the system is shifted passive without considering these constant terms, then the periodic orbit is globally attractive. To the best of our knowledge, this is the first global stability result for non-nominal steady states of the microgrid in the full state space, which provides new insights into the synchronization phenomenon where the dissipativity of the system ensures convergence. The proposed controller is verified with a test microgrid, demonstrating its stability and transient smoothness compared to the standard droop control.
\end{abstract}

\end{frontmatter}

\section{Introduction} \label{sec_intro}
Microgrids are considered key to achieving the sustainability of power systems in their transition from centralized generation to distributed energy resources (DER). Microgrid primary control refers to the lower-level control that defines the transient dynamics under sudden load or generation changes. Besides its steady-state characteristics, the primary control of the microgrid is designed to maintain stability under large power disturbances without relying on communication between DER units~\cite{olivares2014trends}. Hence, stability analysis of the microgrid primary control is the basis for microgrid functionalities, where the main problem is to find the optimal DER control scheme and parameters to ensure convergence of the overall microgrid to a steady state through a sufficiently smooth transient.

A prerequisite for the primary control to maintain stability is to allow some flexibility in the DER power output, so that random power disturbances can be shared when steady state is reached. Steady-state power sharing is usually provided by the droop characteristic of the control (droop control), where active and reactive power outputs are lowered in response to the increases in voltage frequency and amplitude. Various implementations of droop control can be roughly divided into the swing equation~\cite{schiffer2014conditions,de2017bregman,schiffer2019global,8454492} and virtual oscillators~\cite{lu2022virtual,subotic2020lyapunov,he2023nonlinear}. In analyzing its stability, the variable steady-state frequency creates significant challenges, which can be explained by the synchronized steady state being a limit cycle~\cite{garofalo2016energy,yi2020orbital}. \textcolor{myblue}{The frequency of the limit cycle is associated with the power sharing in the microgrid and is not fixed. Hence, the control objective is different from the usual a set point tracking problem as studied in~\cite{riverso2014plug,tucci2020scalable,strehle2021unified}.}

The limit cycle steady state of the microgrid is related to the angle symmetry of the system, i.e., changing the angle reference leaves the dynamics unchanged. We refer to this type of steady state as a circular limit cycle (Section~\ref{sec_prelim}). The usual approach to study the stability of circular limit cycles is to remove the symmetry by changing to transverse coordinates, i.e., replacing absolute angles by angle differences, and thereby reducing the limit cycle to an equilibrium point~\cite{kolluri2018stability,budanur2015periodic,rowley2003reduction}. However, the following limitations exist with this approach: 
\begin{enumerate}
\item[1)] \textcolor{myblue}{In transverse coordinates, the sign-definite energy dissipation no longer stays sign-definite where it is mixed with the power sources. On the other hand, in the original coordinates, the power sources, once integrated, lead to monotonically decaying energy functions that are not lower bounded~\cite{schiffer2014conditions,de2017bregman}. As a compromise, the standard technique is to show that the unbounded energy function, having been transformed into transverse coordinates, is locally convex at the isolated equilibrium point in transverse coordinates~\cite{schiffer2014conditions,de2017bregman}. This certifies it as a local Lyapunov function. However, this technique cannot be used to obtain global stability results.}
\item[2)] \textcolor{myblue}{In transverse coordinates where the voltage angle dynamics are considered, the equilibrium points of the angle variables are isolated but repeat periodically in the space of $\mathbb{R}^n$. The fundamental problem is the periodicity of the original state space of $\mathbb{T}^n$~\cite{forni2014differential}. This problem is overcome in~\cite{schiffer2019global}, which first defines a Leonov function to bound the voltage angles in $\mathbb{R}^n$ before applying the invariance principle to obtain the global convergence result.}
\item[3)] 
\textcolor{myblue}{Due to the complexities above, almost all existing studies only consider simplified models of the system, assuming time-scale separation. Attempts to amend these results to include the electromagnetic network dynamics lead to conservatism~\cite{subotic2020lyapunov}.}
\end{enumerate}

To date, few studies have addressed the stability of microgrids as the stability of a limit cycle (orbital stability) in the original coordinates. Some existing methods for studying orbital stability include Floquet theory and Poincar\'e map~\cite{perko2013differential}, method of slices~\cite{budanur2015periodic}, index iteration theory~\cite{duan2020index,ekeland2012convexity}, and transverse contraction~\cite{manchester2014transverse}. Existing Lyapunov analyses of orbital stability rely on finding transverse and parallel coordinates~\cite{underactuated,hauser1994converse,yi2020orbital,manchester2014transverse}. 
Other than transverse contraction, these methods suffer from the same issue with formulating dissipation in alternative coordinates, \textcolor{myblue}{and transverse contraction may be too conservative for large-scale network systems.} Recently, there have been several orbital stability studies for the dispatchable virtual oscillator control (dVOC). For example, \textcolor{myblue}{\cite{subotic2020lyapunov} constructs a Lyapunov function for the full-order dynamics from that of the globally asymptotically stable reference dynamics by formalizing the time-scale separation assumption;\cite{he2023nonlinear} proposes a stability condition for a non-nominal steady state of dVOC such that the available dissipation in complex coordinates dominates the amplitude-regulating terms.}
Among these existing studies, an intrinsic approach to the orbital stability of microgrids in terms of the angle symmetry and the different reference frames from which the limit cycle originates is still missing.

In this paper, we propose a primary DER control and a stability analysis for the non-nominal limit cycle steady states of the controlled microgrid. Firstly, the angle symmetry of the microgrid system is established. Then, the proposed control, which can be seen as a second-order integrator in the stationary reference frame, is presented.
Assuming the existence of a circular limit cycle, we formulate the system as a port-Hamiltonian system with an artificial shrinking imposed on the solution sets. The latter allows us to decouple the voltage controller from the swing equation in the model. The energy balance of the shifted storage function is carefully derived so that the sign-indefinite terms related to the angle-dependent power sources are eliminated by shifting the phase of the steady state. Then, through analysis of the limiting solution, the circular periodic orbit of the microgrid is proved to be globally attractive under the shifted passivity condition. 
Finally, the control gain tuning is formulated based on the shifted passivity condition for DERs. The contributions are summarized as follows.

\begin{itemize}
    \item A grid-forming DER control is proposed to ensure the global stability of the non-nominal microgrid steady states, independently of the parameters of the swing equation or the network topology that determine power sharing.
    \item The design of the DER control gain is formulated as a semidefinite program for each DER with only local parameters, which provides a distributed and robust stability guarantee for the whole microgrid.
    \item The concept of shifted passivity, suitable for verifying the stability of nonzero equilibrium points, is extended to the synchronization problem for phase oscillators with dynamic passive coupling, to connect passivity with the attractivity of the periodic orbit resulting from synchronization.
\end{itemize}

This paper is structured as follows. Basic definitions are given in Section~\ref{sec_prelim}. The model of the microgrid with the proposed control is presented in Section~\ref{sec_mg}. The port-Hamiltonian model of the DER subsystem is constructed in Section~\ref{sec_PH}, and the stability proof is presented in Section~\ref{sec_eb}. The tuning of the control gain is in Section~\ref{sec_control_param}. Numerical example and conclusion are in Sections~\ref{sec_numerical} and \ref{sec_conclusion}.

{\bf Notation.}
The imaginary unit is $j$. The matrix form of $j$ is $\matr J = \big[ 0, -1;1, 0\big]$. 
A zero vector is $0_n$; a vector of ones is $1_n$; an identity matrix is $\matr I_n$; and a zero matrix is $\matr 0_{n\times m}$. 
The standard basis for $\mathbb{R}^n$ is $\{\mathbf e_1,\ldots, \mathbf e_n\}$.
For an $\matr A \in \mathbb{C}^{n\times n}$, the transpose is $\matr A^\tran$, 
the Hermitian transpose is $\matr A^*$, \textcolor{myblue}{the Moore-Penrose pseudoinverse is $\matr A^\dagger$,
and the Hermitian part is $\mathrm{He}\{\matr A\} = \frac{1}{2}(\matr A + \matr A^*)$.}
Denote by $\mathrm{diag}(\matr A_1,\ldots,\, \matr A_n)$ a block diagonal matrix with $\matr A_i,\, i = 1,\ldots,\, n$ on the diagonal. 
Denote by $\mathrm{col}(\vect x_i)$ a column vector that stacks the vectors $\vect x_i,\, i = 1,\ldots,\, n$. 
The standard inner product for $\mathbb{C}^n$ is $\langle\vect y, \vect x\rangle = \Re\{\vect y^* \vect x\}$. 
For a Hermitian matrix $\matr A$, positive definiteness is denoted by $\matr A \succ 0$, i.e., $\langle \vect x, \matr A \vect x\rangle > 0$ for all $\vect x \in \mathbb{C}^n \neq 0_n$. 
\textcolor{myblue}{For an $\matr A \succeq 0$, denote $\|\vect x\|_{\matr A}^2 = \langle \vect x, \matr A \vect x \rangle$.} For a function $g: \mathbb{C}^n \to \mathbb{R}$ of complex variables, the complex gradient is $\nabla g(\vect x) = 2 \mathrm{col}(\frac{\partial g}{\partial x_i^*})$ where $\frac{\partial g}{\partial x_i^*}$ is the Wirtinger derivative~\cite{remmert1991theory}.
\textcolor{myblue}{Nominal set points for the control are denoted by $(\cdot)_\mathsf{nom}$. The Kronecker product is $\otimes$.}

\section{Preliminaries} \label{sec_prelim}

\subsection{Symmetry and circular limit cycle}

The circular limit cycle is defined in this subsection. A formal treatment of symmetries of ODEs can be found in Chapter~1 of~\cite{bocharov1999symmetries}. The group characterization can be found in~\cite{kozlov1988symmetry}.

\begin{defn} \label{def_one_para}
A one-parameter group is a differentiable homomorphism $\mathcal T: \mathbb{R}\to G$ from $\mathbb{R}$ (additive group) to a topological group $G$ such that $\mathcal T(\tau) \mathcal T(s) = \mathcal T(\tau + s)$. The infinitesimal generator of $\mathcal T(\tau)$ is $\mathcal G = \frac{\partial\mathcal T(\tau)}{\partial\tau}\big|_{\tau = 0}$.
\end{defn}

\begin{defn} \label{def_sym}
For the system $\dot{\vect x} = \vect f(\vect x)$ where $\vect x \in \mathbb{C}^n$,
a continuous symmetry is a one-parameter group of automorphisms $\mathcal T(\tau): \mathbb{C}^n \to \mathbb{C}^n$, such that each $\mathcal T(\tau)$ maps solutions of the system to solutions.
\end{defn}

The microgrid system considered in this paper has the angle symmetry: $\mathcal T(\tau)\vect x = e^{j\tau}\vect x$. Since this $\mathcal T(\tau)$ is $2\pi$-periodic in the parameter $\tau$, $\tau$ naturally lives on the $1$-torus $\mathbb{T} = \mathbb{R}/2\pi \mathbb{Z}$ with $2\pi$ equivalence~\cite{warner1983foundations}.

Let us consider the implications of angle symmetry on the possible limiting solutions of the system.\footnote{A limiting solution is a solution whose orbit is contained in a limit set~\cite{lasalle1960some}.}
For a given limiting solution $\eqm{\vect x}(t)$, the following two scenarios exist:
\begin{enumerate}
    \item[i)] There is an $\eqm\omega \in \mathbb{R} \neq 0$ such that $\dot{\eqm{\vect x}}(t) = \eqm\omega \mathcal G \eqm{\vect x}(t) = j \eqm\omega \eqm{\vect x}(t)$ for all $t \in \mathbb{R}$. This is the case considered in this paper.
    \item[ii)] There is no such $\eqm\omega$. The limit set is not isolated in this case according to Definition~\ref{def_sym}.
\end{enumerate}
 
For {i)}, the limiting solution is a limit cycle given by
\begin{equation} \label{E:limit_cycle}
    \vect x(t) = e^{j\eqm\omega t} \vect x(0),
\end{equation}
where $\eqm\omega$ is the frequency of the limit cycle.
From Definition~\ref{def_sym}, mappings of $\vect x(t)$ in (\ref{E:limit_cycle}) by $\mathcal T(\tau)$ are also limit cycle solutions---with shifted phases. Let us designate  one of these limit cycle solutions $\hat{\vect x}(t)$ to have phase $0$; this set of limit cycle solutions can be denoted as
\begin{equation*}
    \eqm{\vect x}(\tau, t) = \mathcal T(\tau)\hat{\vect x}(t),
\end{equation*}
for $\tau \in \mathbb{T}$,
all of which share the circular orbit,
\begin{equation} \label{E:periodic_orbit}
    \mathrm{span}(\eqm{\vect x}(0, t)) = \big\{ e^{j\tau} \hat{\vect x}(0) \mid \tau \in \mathbb{T} \big\}.
\end{equation}
In the sequel, the time parameter in $\eqm{\vect x}(\tau, t)$ is omitted, resulting in the notation $\eqm{\vect x}(\tau)$ for a limit cycle solution of phase $\tau$ with the periodic orbit $\mathrm{span}(\eqm{\vect x}(\tau))$.

If $\eqm\omega \neq 0$,\footnote{The degenerate case $\eqm\omega = 0$ can be removed by changing to a reference frame rotating at some nonzero frequency.} the limit cycle in scenario {ii)} is identified by the the following condition,
\begin{equation} \label{E:lc_condition}
    (\exists\, \vect x \in \mathbb{C}^n,\, \eqm\omega \in \mathbb{R}\neq 0) \text{ s.t. } \vect f(e^{j\tau} \vect x) = j \eqm\omega e^{j\tau} \vect x,\; \forall \tau \in \mathbb{T}.
\end{equation}
Due to the additional parameter $\eqm\omega$, the condition (\ref{E:lc_condition}) for circular limit cycles is slightly harder to solve than the equilibrium point condition. To avoid this problem, it is commonplace in the literature to assume that $\eqm\omega$ is equal to the nominal $60$ (or $50$) Hz. Then, in a reference frame rotating at frequency $\eqm\omega$, the dynamics is transformed into $\dot{\vect x} =  - j\eqm\omega \vect x + \vect f(\vect x)$, for which the equilibrium condition is solved. \textcolor{myblue}{However, in practice, the steady-state frequency $\eqm\omega$ is almost never equal to the nominal frequency due to the complex balance between power supply and  demand~\cite{gross2018steady}. In other words, an equilibrium point usually cannot be found in the nominal frequency reference for the precise model. This motivates the stability analysis for non-nominal limit cycles~\cite{he2023nonlinear}. }

\subsection{Shifted passivity of port-Hamiltonian system}

The shifted passivity conditions proposed in~\cite{monshizadeh2019conditions} serve to characterize the passivity of port-Hamiltonian (pH) systems with respect to a nonzero equilibrium point. In this subsection, we introduce pH systems in the state space $\mathbb{C}^n$, which is a mild extension to the $\mathbb{R}^n$ case in~\cite{van2014port}, and recall basic results on shifted passivity.

Consider a generic pH model for a subsystem of a network, of the form~\cite{van2014port}
\begin{subequations} \label{E:PH} \begin{align}
    \dot{\vect x} &= \matr F \nabla H(\vect x) + \matr G \vect u \\
    \vect y &= \matr G^* \nabla H(\vect x),
\end{align} \end{subequations}
where $\vect x \in \mathbb{C}^n$ is the state vector, $H: \mathbb{C}^n \to \mathbb{R}_{\geq 0}$ is the Hamiltonian, 
$\matr F \in \mathbb{C}^{n\times n}$ is the system matrix, and $\matr G = \big[ \matr G_1,\, \matr G_2 \big]$ is the input matrix. The input $\vect u = \big[\vect u_1^\tran,\, \vect u_2^\tran \big]^\tran$ and output $\vect y = \big[ \vect y_1^\tran,\, \vect y_2^\tran \big]^\tran$ are grouped as follows: 
\begin{enumerate}
    \item[1)] $(\vect u_1, \vect y_1)$ represents interactions within a microgrid network system being studied,
    \item[2)] $(\vect u_2, \vect y_2)$ represents external disturbances such as load and mechanical input.
\end{enumerate} 
We denote by $\Omega \ni (\vect x(t),\, \vect u(t),\, \vect y(t))$ the solution set of the pH system (\ref{E:PH}) for any input function $\vect u(t)$.

For a continuously differentiable function $H: \mathbb{C}^n \to \mathbb{R}$, let us define the shifted version of $H(\vect x)$ centered at an $\eqm{\vect x} \in \mathbb{C}^n$ as
\begin{equation} \label{E:shifted_function}
    \mathcal H(\vect x, \eqm{\vect x}) = H(\vect x) - H(\eqm{\vect x}) - \langle \nabla H(\eqm{\vect x}), \vect x - \eqm{\vect x}\rangle,
    \vspace{4pt}
\end{equation}
which is the storage function for verifying shifted passivity~\cite{monshizadeh2019conditions,de2017bregman}. 
Its gradient w.r.t. $\vect x$ is
\begin{equation} \label{E:grad_H}
    \nabla_{\vect x} \mathcal H(\vect x, \eqm{\vect x}) = \nabla H(\vect x) - \nabla H(\eqm{\vect x}).
\end{equation}
\emph{Claim:}  If $H(\vect x)$ is quadratic, i.e., $H(\vect x) = \frac{1}{2} \langle \vect x, \matr Q \vect x \rangle$ for some $\matr Q = \matr Q^*$, then 
\begin{equation} \label{E:s1-2}
    \nabla_{\eqm{\vect x}} \mathcal H(\vect x, \eqm{\vect x}) = -\nabla_{\vect x} \mathcal H(\vect x, \eqm{\vect x}).
\end{equation}
From (\ref{E:shifted_function}), we get $\mathcal H(\vect x, \eqm{\vect x}) = \frac{1}{2} \langle \vect x - \eqm{\vect x}, \matr Q (\vect x - \eqm{\vect x}) \rangle$. Since $\mathcal H(\vect x, \eqm{\vect x})$ is a function of $\vect x - \eqm{\vect x}$, we obtain (\ref{E:s1-2}).

\begin{defn} \label{def_sp}
Consider the pH system (\ref{E:PH}) and an equilibrium point $(\eqm{\vect x}, \eqm{\vect u})$. Let $\eqm{\vect y} = \matr G^* \eqm{\vect u}$.
The pH system is said to be shifted passive w.r.t. $(\eqm{\vect x}, \eqm{\vect u})$ if, for all $\vect x$, it holds that
\begin{equation*}
    \dot{\mathcal H}(\vect x, \eqm{\vect x}) \leq \langle \vect y- \eqm{\vect y}, \vect u - \eqm{\vect u} \rangle.
\end{equation*}
\end{defn}

To interpret Definition~\ref{def_sp}, let us write the energy balance as
\begin{align}
    &\dot{\mathcal H}(\vect x, \eqm{\vect x}) = \langle \nabla_{\vect x} \mathcal H(\vect x, \eqm{\vect x}), \dot{\vect x}\rangle = \langle \nabla_{\vect x} \mathcal H(\vect x, \eqm{\vect x}), \dot{\vect x} - \dot{\eqm{\vect x}} \rangle \label{E:s1-1} \\
    &= \langle \nabla H(\vect x) - \nabla H(\eqm{\vect x}), \matr F [\nabla H(\vect x) - \nabla H(\eqm{\vect x})] + \matr G (\vect u - \eqm{\vect u}) \rangle \notag \\
    &= \langle \nabla H(\vect x) - \nabla H(\eqm{\vect x}), \matr F [\nabla H(\vect x) - \nabla H(\eqm{\vect x})] \rangle \notag \\
    &\quad\, + \langle \vect y - \eqm{\vect y}, \vect u - \eqm{\vect u} \rangle, \label{E:energy_balance_pre}
\end{align}
where we subtracted $\dot{\eqm{\vect x}} = 0_n$ in (\ref{E:s1-1}) and applied the adjoint of $\matr G$ in (\ref{E:energy_balance_pre}).
From (\ref{E:energy_balance_pre}), the shifted passivity of (\ref{E:PH}) holds as soon as $\mathrm{He}\{\matr F\} \prec 0$. Note that if the Hamiltonian $H(\vect x)$ is quadratic, then due to (\ref{E:s1-2}), (\ref{E:s1-1}), the energy balance equation (\ref{E:energy_balance_pre}) still holds if $\eqm{\vect x}$ is not an equilibrium point, i.e., $\dot{\eqm{\vect x}} \neq 0_n$. This fact is exploited to extend shifted passivity to limit cycles later. 
Lastly, for memoryless mappings, we have the following definition.

\begin{defn}
The input-output mapping $\Upsilon: \mathbb{C}^m\to \mathbb{C}^m$ is said to be shifted passive w.r.t. the steady-state input $\eqm{\vect u}$ if for all $\vect u$, it holds that
\begin{equation} \label{E:memoryless}
    0 \leq \langle \Upsilon(\vect u) - \Upsilon(\eqm{\vect u}), \vect u - \eqm{\vect u}\rangle.
\end{equation}
\end{defn}

If (\ref{E:memoryless}) holds for all $\eqm{\vect u} \in \mathbb{C}^n$, this property is also known as equilibrium independent passivity~\cite{simpson2018equilibrium}, and as monotonicity~\cite{ryu2016primer}.

\section{Model of the microgrid system with DER control} \label{sec_mg}
In this section, we model the microgrid using the hierarchical port-Hamiltonian approach in~\cite{fiaz2013port}. 
The system is viewed as a directed graph, where the edges include the DER generator ($G$), $R$--$L$ line ($T$), and shunt capacitor ($C$) edges, assuming the $\Pi$-model of the distribution lines~\cite{watson2021scalable}. Different from~\cite{fiaz2013port}, we treat the loads, assumed static~\cite{strehle2021unified}, as additive disturbances to the shunt capacitor dynamics. The nodes include the DER generator, \textcolor{myblue}{standalone load}, and ground node. To simplify the model, we assume i) the ground node has zero potential; ii) the $R$--$L$ lines have no interior junctions. 
Let the microgrid comprise $\textup{\texttt{g}}$ DER generators, $\textup{\texttt{T}}$ $R$--$L$ lines, \textcolor{myblue}{and $\textup{\texttt{g}} + \ell$ shunt capacitors with loads that are connected from every node to the ground}. The total number of edges is $\mathtt n = 2\textup{\texttt{g}} + \ell + \textup{\texttt{T}}$, with the indices $\textup{\texttt{e}} = 1,\ldots, \mathtt n$.\footnote{To avoid confusion with other subscripts, the subscript $\textup{\texttt{e}}$ is never replaced by its value.}
The edges are ordered as follows.
Edges $1$ to $\textup{\texttt{g}}$ are DER generators. Edges $\textup{\texttt{g}} + 1$ to $\textup{\texttt{g}} + \textup{\texttt{T}}$ are $R$--$L$ lines. \textcolor{myblue}{The remaining $\texttt g + \ell$ edges are shunt capacitors with loads.} 
The edge voltage and current directions follow the direction of the edge such that the ground is always the head.  
The incidence matrix for the graph of the power network is given by
\begin{equation*}
    \matr M = \begin{bmatrix}
        \begin{bmatrix}
            \matr I_{\textup{\texttt{g}}} \\
            \matr 0_{\ell\times \textup{\texttt{g}}}
        \end{bmatrix} &\matr M_1 &\matr I_{\textup{\texttt{g}} + \ell}  \\
        -1_{\textup{\texttt{g}}}^\tran &0_{\textup{\texttt{T}}}^\tran &-1_{\textup{\texttt{g}} + \ell}^\tran 
    \end{bmatrix}
\end{equation*}
where $\matr M_1$ is the incidence matrix of the sub-graph obtained by removing the ground node~\cite{fiaz2013port}.

\subsection{Interconnection constraints}

We assume that the microgrid is balanced, i.e., the sum of every three-phase voltage or current is zero. Then, applying the power-preserving Clarke ($\alpha\beta0$ or stationary $dq0$) transformation~\cite{o2019geometric} to every three-phase voltage or current $\vect x_{abc} = \big[x_a,\, x_b,\, x_c \big]^\tran$, the $0$-component evaluates to zero, and the $\alpha\beta$-components are represented by a single complex variable $x_{\alpha\beta} = x_\alpha + j x_\beta$. The mapping $\vect x_{abc} \mapsto x_{\alpha\beta}$ is linear, unitary, and both $\vect x_{abc}$ and $x_{\alpha\beta}$ have two degrees of freedom.

As introduced later, the dynamics of each edge has the pH form (\ref{E:PH}), which are interconnected by the constraints on the port variables $(\vect u_{\texttt e, 1}, \vect y_{\texttt e, 1})$.
Let us denote by $\underline U = \big[\underline V_G^\tran,\, \underline V_T^\tran,\, \underline V_{C}^\tran\big]^\tran$ the edge voltages, $\underline I = \big[\underline I_G^\tran,\, \underline I_T^\tran,\, \underline I_{C}^\tran\big]^\tran$ the edge currents, and $\underline V = \big[\underline V_C^\tran,\, 0\big]^\tran$ the node potentials.
The physical constraints are given by
\begin{align*}
    &\matr M \underline I = 0_{\texttt n}\; \text{(KCL)},\\
    &\matr M^\tran \underline V = \underline U\; \text{(KVL).}
\end{align*}
For DER generators and $R$--$L$ lines, we assign the edge current as the input $\vect u_{\texttt e, 1}$. For shunt capacitors, we assign the edge voltage as the input $\vect u_{\texttt e, 1}$.\footnote{Complementary inputs and outputs are chosen for interacting edges for the interconnection constraints to be a linear mapping with no additional algebraic constraints.} The edge inputs and outputs are stacked into
\begin{align*}
    \mathrm{col}(\vect u_{\texttt e, 1}) = \big[\underline V_G^\tran,\, \underline V_T^\tran,\, \underline I_{C}^\tran\big]^\tran \\
    \mathrm{col}(\vect y_{\texttt e, 1}) = \big[\underline I_G^\tran,\, \underline I_T^\tran,\, \underline V_{C}^\tran\big]^\tran.
\end{align*}
We then obtain, from KCL and KVL, the interconnection constraints $\mathrm{col}(\vect u_{\texttt e, 1}) = \matr W \mathrm{col}(\vect y_{\texttt e, 1})$
where
\begin{equation} \label{E:edge_constraints}
    \matr W = \begin{bmatrix}
        \matr 0_{\textup{\texttt{g}}\times \textup{\texttt{g}}} &\matr 0_{\textup{\texttt{g}}\times \textup{\texttt{T}}} &\begin{bmatrix}
            \matr I_{\textup{\texttt{g}}} &\matr 0_{\textup{\texttt{g}}\times \ell}
        \end{bmatrix} \\
        \matr 0_{\textup{\texttt{T}}\times \textup{\texttt{g}}} &\matr 0_{\textup{\texttt{T}}\times \textup{\texttt{T}}} &\matr M_1^\tran \\
        \begin{bmatrix}
            -\matr I_{\textup{\texttt{g}}} \\
            \matr 0_{\ell\times \textup{\texttt{g}}}
        \end{bmatrix} &-\matr M_1 &\matr 0_{\textup{\texttt{g}} + \ell}
    \end{bmatrix}.
\end{equation}

\subsection{Edges for $R$--$L$ lines ($\textup{\texttt{e}} = \textup{\texttt{g}} + 1, \ldots,\, \textup{\texttt{g}} + \textup{\texttt{T}}$)} \label{sec_RL}
The equation for the $R$--$L$ line is given by
\begin{equation} \label{E:R-L}
    \dot{(L_\textup{\texttt{e}} I_\textup{\texttt{e}})} = -R_\textup{\texttt{e}} I_\textup{\texttt{e}} + V_\textup{\texttt{e}},
\end{equation}
where $R_{\texttt e}$ and $L_{\texttt e}$ are the resistance and inductance of the $R$--$L$ line.
The subsystem (\ref{E:R-L}) is rewritten in the pH form (\ref{E:PH}) by defining the state $\vect x_\textup{\texttt{e}} = L_\textup{\texttt{e}} I_\textup{\texttt{e}}$, the input $\vect u_{\texttt{e},1} = V_\textup{\texttt{e}}$, the Hamiltonian $H_\textup{\texttt{e}}(\vect x_\textup{\texttt{e}}) = \frac{1}{2} L_{\texttt e}^{-1} \|\vect x_{\texttt e}\|^2$ with $\nabla H_{\texttt e}(\vect x_{\texttt e}) = I_{\texttt e}$, 
$\matr F_{\texttt{e}} = -R_\texttt{e} \matr I_1$, and $\matr G_{\texttt{e}, 1} = \matr I_1$. We can find the output $\vect y_{\texttt{e},1} = \matr G_{\texttt e, 1}^* \nabla H_{\texttt e}(\vect x_{\texttt e}) = I_\textup{\texttt{e}}$.

\subsection{Edges for shunt capacitors with static loads ($\textup{\texttt{e}} = \textup{\texttt{g}} + \textup{\texttt{T}} + 1,\ldots,\, 2\textup{\texttt{g}} + \ell + \textup{\texttt{T}}$)} \label{sec_shunt}

The equation for the shunt capacitor with the load disturbance is given by~\cite{watson2021scalable}
\begin{equation} \label{E:shunt}
    \dot{(C_\textup{\texttt{e}} V_\textup{\texttt{e}})} = -G_{\texttt e} V_{\texttt e} + I_\textup{\texttt{e}} - \Upsilon_\textup{\texttt{e}}(V_\textup{\texttt{e}}),
\end{equation}
where $C_{\texttt e}$ and $G_{\texttt e}$ are the capacitance and conductance of the shunt capacitor, $\Upsilon_{\texttt e}(V_{\texttt e})$ is the load current as a function of the bus voltage. The subsystem  (\ref{E:shunt}) is rewritten in the pH form (\ref{E:PH}) by defining the state $\vect x_\textup{\texttt{e}} = C_\textup{\texttt{e}} V_\textup{\texttt{e}}$, the inputs $\vect u_{\texttt{e}, 1} = I_{\texttt e},\, \vect u_{\texttt e, 2} = -\Upsilon_{\texttt e}(V_{\texttt e})$, the Hamiltonian $H_\textup{\texttt{e}}(\vect x_\textup{\texttt{e}}) = \frac{1}{2} C_\textup{\texttt{e}}^{-1} \|\vect x_\textup{\texttt{e}}\|^2$ with $\nabla H_{\texttt e}(\vect x_{\texttt e}) = V_{\texttt e}$, 
$\matr F_\textup{\texttt{e}}(\vect x_{\texttt e}) = -G_{\texttt e} \matr I_1$, and $\matr G_{\texttt e, 1} = \matr G_{\texttt e, 2} = \matr I_1$. The outputs are $\vect y_{\texttt e, 1} = \vect y_{\texttt e, 2} = V_{\texttt e}$. 

\begin{assum} \label{assum_load_sym}
Assume that the load current function is angle-invariant, i.e., $e^{j\tau} \Upsilon_\textup{\texttt e}(V_\textup{\texttt e}) = \Upsilon_\textup{\texttt e}(e^{j\tau} V_\textup{\texttt e})$.
\end{assum}

\subsection{Edges for DER generators ($\textup{\texttt{e}} = 1,\ldots,\, \textup{\texttt{g}}$)} \label{sec_DER}

In this subsection,\footnote{The subscript $\texttt e$ is dropped in this subsection to ease notation.} equations for the closed-loop DER subsystem with the proposed controller are given. For more background, the reader is referred to~\cite{pogaku2007modeling} for the PI control in the internal reference frame and~\cite{teodorescu2006proportional,pereira2015tuning} for the PR control in a stationary reference frame. 
All variables in this subsection are defined in the stationary reference frame.

I. \emph{Swing equation} with approximate active power:
\begin{align*}
    J \dot{\omega} &= - D (\omega - \omega_{\mathsf{nom}}) - (V_\mathsf{nom} \Re\{I_t e^{-j\theta}\} - 
    P_{\mathsf{nom}})/\omega_\mathsf{nom} \\
    \dot{\theta} &= \omega,
\end{align*}
where $J$ is the inertia, $D$ is the damping constant, $\omega_{\mathsf{nom}}$ and $P_{\mathsf{nom}}$ are the nominal frequency and output active power, and $V_\mathsf{nom} \Re\{I_t e^{-j\theta}\}$ is the approximate output active power such that the nominal voltage is used to calculate the active power instead of the instantaneous voltage.
The precise active power $\Re\{I_t^* V_o\}$ is equal to the approximated active power if $V_o = V_{\mathsf{nom}} e^{j\theta} - j X I_t$, which is achieved when the voltage controller below reaches steady state. In complex coordinates (\ref{E:1}) is expressed as
\begin{subequations} \label{E:1} \begin{align}
    J \dot{(j \omega e^{j\theta})} &= -J \omega^2 e^{j\theta} - D (j \omega e^{j\theta} - j \omega_{\mathsf{nom}} e^{j\theta}) \notag \\
    &\quad\, - j e^{j\theta} (V_\mathsf{nom} \Re\{I_t e^{-j\theta}\} - 
    P_{\mathsf{nom}})/\omega_\mathsf{nom} \label{E:swing_equation} \\
    \dot{(e^{j\theta})} &= j \omega e^{j\theta}. \label{E:swing_angle}
\end{align} \end{subequations}

II. The proposed \emph{voltage controller:}
\begin{subequations} \label{E:2} \begin{align}
    \dot{\beta} &= V_{\mathsf{nom}} e^{j\theta} - j X I_t - V_o \label{E:voltage_loop} \\
    \dot{\xi} &= j\omega_\mathsf{nom} \xi + K_{iv} \beta \label{E:current_loop} \\
    m &= \matr K \Big[I_b,\, V_o,\, I_t,\, \beta,\, \xi \Big]^\tran, \label{E:mod_index}
\end{align} \end{subequations}
where $\beta$ and $\xi$ are the state variables. 
The parameters include the virtual reactance $X$~\cite{wang2014virtual}, the gain $K_{iv}$, and $\matr K \in \mathbb{C}^{1\times 5}$. The output signal $m$ is the modulation index for the inverter to create the terminal voltage $V_t = \frac{V_{dc}}{2} m$ (ignoring the switching frequency).
Intuitively, the first integrator (\ref{E:voltage_loop}) achieves $\eqm V_o(\tau) \approx V_{\mathsf{nom}} e^{j\eqm\theta(\tau)} - j X \eqm I_t(\tau)$, and the second integrator (\ref{E:current_loop}) achieves $\eqm\beta(\tau) \approx 0$. By substituting (\ref{E:voltage_loop}) and (\ref{E:current_loop}) into the limit cycle condition, $\dot{\eqm{\vect x}}(\tau) = j\eqm\omega \eqm{\vect x}(\tau)$, we obtain the steady-state voltage as:
\begin{align} 
    \eqm V_o(\tau) &= V_{\mathsf{nom}} e^{j\eqm\theta(\tau)}  - j X \eqm I_t(\tau) + \frac{\eqm\omega (\eqm\omega - \omega_{\mathsf{nom}})}{K_{iv}} \eqm\xi(\tau). \label{E:voltage_error}
\end{align} 

\usetikzlibrary{arrows.meta}
\ctikzset{bipoles/vsourceam/inner plus={\tiny $+$}}
\ctikzset{bipoles/vsourceam/inner minus={\tiny $-$}}

\begin{figure}[!t]
\hspace{-0.12in}
\scalebox{0.95}{
\begin{circuitikz}[american]
    \ctikzset{
        resistors/scale=0.4,
        capacitors/scale=0.4,
        inductors/scale=0.4,
        sources/scale=0.6
    }
    
    \node[draw, minimum width=1.1cm, minimum height=1.6cm, line width=0.7pt] (VSC) at (-0.6,-0.6) {};
    \node[Lnigbt, bodydiode, left=0.52cm, scale=0.55] at (VSC.east){};
    \draw[color=blue!95] (3.2,0) ellipse (0.08cm and 0.12cm);
    \draw[color=blue!95] (3.7,0) ellipse (0.08cm and 0.12cm);
    \node[draw, line width=0.7pt, text width=1.8cm, align=center, color=blue!95] (control) at (-0.5,1.2) {\scriptsize Second-order \\ int. (\ref*{E:2})};
    \node[draw, text width=1.4cm, line width=0.7pt, align=center, color=blue!95] (swing) at (2.0,1.2) {\scriptsize Swing eq.\\ (\ref*{E:1})};
    \node[draw, circle, minimum size=0.3cm, line width=0.7pt, color=blue!95] (sum) at (0,3.0){};
    \node[draw, minimum width=0.5cm, minimum height=0.3cm, line width=0.7pt, color=blue!95] (virtual) at (2.55,2.5) {\scriptsize $jX$};
    \node[draw, minimum width=0.5cm, minimum height=0.3cm, line width=0.7pt, color=blue!95] (Vn) at (1.1,3.0) {\scriptsize $V_{\mathsf{nom}}$};

    \ctikzset{flow/distance=0cm}
    \draw (-0.05,0) -- (0.8,0) to[R,l^=\scriptsize $R_f$] (1.8,0) to[L,l^=\scriptsize $L_f$,f_=\scriptsize $I_f$,current arrow scale=20] (2.3,0) -- (4.2,0) to[R,l^=\scriptsize $R_c$] (5.2,0) to[L,l^=\scriptsize $L_c$,f_=\scriptsize $I_b$,current arrow scale=20] (5.7,0) -- (6.3,0) node[circ, scale=0.7]{};
    \draw (3.7,0) node[circ, scale=0.7]{} to[C,l_=\scriptsize $C_f$] (3.7,-1.2) node[circ, scale=0.7]{};
    \draw (3.7,-1.2) node[ground] {};
    \draw (-0.05,-1.2) -- (6.3,-1.2) node[circ, scale=0.7]{};
    \draw[-{Latex[round]},color=blue!95] (3.7,0.12) |- (0,3.6) node[label={[xshift=0.35cm, yshift=-0.18cm]\scriptsize $V_o$}]{} -- (sum.north) node[left=5pt, above]{\tiny $-$};
    \draw[-{Latex[round]},color=blue!95] (3.2,1.2) node[circ, scale=0.7]{} |- (virtual.east) node[label={[xshift=0.45cm, yshift=-0.18cm]\scriptsize $I_t$}]{};
    \draw[color=blue!95] (virtual.west) -- (2.08,2.5);
    \draw[-{Latex[round]},color=blue!95] (1.92,2.5)-| (sum.south) node[left=5pt,below]{\tiny $-$};
    \draw[color=blue!95] (2.08,2.5) arc (0:180:0.08);
    \draw[-{Latex[round]},color=blue!95] (3.2,0.12) |- (swing.east);
    \draw[-{Latex[round]},color=blue!95] (swing.north) |- (Vn.east) node[label={[xshift=0.4cm, yshift=-0.18cm]\scriptsize $e^{j\theta}$}]{};
    \draw[-{Latex[round]},color=blue!95] (Vn.west) |- (sum.east) node[right=5pt, above]{\tiny $+$};
    \draw[-{Latex[round]},color=blue!95] (sum.west) -| (control.north);
    \draw[color=blue!95] ([yshift=55]3.2,0.08) node[circ, scale=0.7]{} -- ([yshift=55]2.08,0.08);
    
    \draw[-{Latex[round]},color=blue!95] ([yshift=55]1.92,0.08)-| ([xshift=10]control.north);
    \draw[color=blue!95] ([yshift=55]2.08,0.08) arc (0:180:0.08);
    \draw[-{Latex[round]},color=blue!95] (0,3.6) node[circ, scale=0.7]{} -| ([xshift=-10]control.north);
    
    \draw[-{Latex[round]},color=blue!95] ([xshift=-10]control.south) -| ([xshift=-7]VSC.north) node[label={[xshift=0.4cm, yshift=-0.08cm]\scriptsize $m$}] {};
    \draw (0.4,-0.05) to[open,v=\scriptsize $V_t$] ++(0,-1.15);
    \draw (4.3,-0.05) to[open,v=\scriptsize $V_o$] ++(0,-1.15);
    \draw (6.5,-0.05) to[open,v=\scriptsize $V_b$] ++(0,-1.15);
    \draw (-1.15,-1.2) -- (-1.6,-1.2) to[american voltage source] (-1.6,0) node[label={[xshift=-0.3cm, yshift=-0.55cm]\scriptsize $V_{dc}$}] {} -- (-1.15,0);
\end{circuitikz}}
\caption{Single-phase diagram of DER control system in a stationary reference frame}
\label{fig_block_diagram}
\end{figure}

III. The $RLC$ filter and the line coupling:
\begin{subequations} \label{E:3} \begin{align}
    L_c \dot I_b &= - R_c I_b + V_o - V_b \label{E:I_b} \\
    C_f \dot V_o &= I_t - I_b \\
    L_f \dot I_t &= - R_f I_t + \frac{1}{2} V_{dc} m - V_o, \label{E:I_t}
\end{align} \end{subequations}
where $V_t = \frac{1}{2} V_{dc} m$ is the inverter terminal voltage. The input and output of the DER subsystem that can interact with the network are $\vect u_1 = V_b$ and $\vect y_1 = -I_b$.

\subsection{Angle symmetry of the microgrid system}

\begin{lem} \label{lem_sym}
Consider the microgrid system consisting of the interconnection constraints (\ref{E:edge_constraints}) and the edge dynamics (\ref{E:R-L}), (\ref{E:shunt}), (\ref{E:1}), (\ref{E:2}), (\ref{E:3}). Let Assumption~\ref{assum_load_sym} hold.
Then this system has the symmetry $\mathcal T(\tau) \vect x = e^{j\tau} \vect x$.
\end{lem}

\begin{pf}
By Definition~\ref{def_sym}, it is equivalent to prove that the solution set of the microgrid system is invariant to $\mathcal T(\tau)$.
We can construct the solution set $\Omega$ of the microgrid through the following two steps: i) Collect the solution sets of every edge,
\begin{equation*}
    \Omega_{\texttt e} \ni (\vect x_{\texttt e}(t),\, \vect y_{\texttt e}(t),\, \vect u_{\texttt e}(t)),
\end{equation*}
where the unconstrained edge inputs include all continuous functions.
It is straightforward to check that every edge equation satisfies $\vect f_{\texttt e}(e^{j\tau} \vect x_{\texttt e}, e^{j\tau} \vect u_{\texttt e}) = e^{j\tau} \vect x_{\texttt e}$. The outputs being linear in $\vect x_{\texttt e}$ are shifted by $e^{j\tau}$ as well.
ii) Keep the unconstrained edge solutions that satisfy the network and the load constraints, i.e.,
\begin{align*}
    &\Omega = \big\{\mathrm{col}(\vect x_{\texttt e}(t)) \mid \mathrm{col}(\vect u_{\texttt e}(t)) = \matr W \mathrm{col}(\vect y_{\texttt e}(t)), \\
    &\hspace{0.1in}\vect u_{\texttt e}(t) = \Upsilon_{\texttt e}(\vect y_{\texttt e}(t)),\, \texttt e = \texttt g + \texttt T + 1,\ldots,\, 2\texttt g + \ell + \texttt T\big\}.
\end{align*}
Since the interconnection constraints and the load constraints are invariant to the angle symmetry, the overall solution set
$\Omega$ is invariant to the angle symmetry. \hfill $\square$
\end{pf}

\section{Port-Hamiltonian model of the DER subsystem} \label{sec_PH}

\begin{assum} \label{assum_lim}
Assume that the microgrid system has a circular limit cycle $\eqm{\vect x}(\tau, t) = e^{j\eqm\omega (t - t_0)} \eqm{\vect x}(\tau, t_0)$.
\end{assum}

In the stability proof, we first consider a modified solution set of the microgrid system where every solution is multiplied by $e^{-\kappa(t - t_0)}$ for $0 < \kappa \ll 1$. The purpose is to include a tiny amount of quadratic dissipation in the complex DER angles (\ref{E:1}b). The equations of the microgrid system are changed from $\dot{\vect x} = \vect f(\vect x, \vect u_2)$ to $\dot{\vect x} = -\kappa \vect x + \vect f(\vect x, \vect u_2')$, where $\vect u_2'(\vect x) = e^{-\kappa(t - t_0)} \vect u_2(e^{\kappa(t-t_0)} \vect x)$.

We decouple the voltage controller (\ref{E:2}) from the swing equation (\ref{E:1}) by incorporating an additional state variable $\varphi$ into (\ref{E:2}):\footnote{For simplicity, we refrain from creating a different notation for the shrinking states. The latter are assumed until Proposition~\ref{prop_main}, where we remove the effect of shrinking in (\ref{E:ineq7}).}
\begin{subequations} \label{E:perturbed} \begin{align}
    \dot{\beta} &= -\kappa \beta - \kappa \varphi - j X I_t - V_o \\
    \dot{\varphi} &= j\omega \varphi - \kappa \varphi \\
    \dot{\xi} &= -\kappa \xi + j\omega_\mathsf{nom} \xi + K_{iv} \beta \\
    m &= \matr K \Big[I_b,\, V_o,\, I_t,\, \beta,\, \xi \Big]^\tran. 
\end{align} \end{subequations}
Let us fix $\varphi(t_0) = \eqm\varphi(\tau, t_0) = -V_\mathsf{nom}/\kappa$. Then, the magnitude $|\varphi| = |\eqm\varphi(\tau)|$ is determined for all time from (\ref{E:perturbed}b), and we have
\begin{equation} \label{E:shrink}
    {-}\kappa \varphi = e^{-\kappa (t - t_0)} V_\mathsf{nom} e^{j\theta},
\end{equation}
which is the shrinking internal reference voltage; similarly for the steady-state solution.

To express the DER subsystem in pH form, we replace $\beta$ and $\varphi$ by $\beta_1 = \beta + \varphi$ and $\beta_2 = \beta - \varphi$ to transform (\ref{E:perturbed}) into
\begin{align}
    \dot \beta_1 &= j \frac{\omega}{2} (\beta_1 - \beta_2) - j X I_t - V_o - \frac{3}{2} \kappa \beta_1 + \frac{1}{2} \kappa \beta_2 \notag \\
    \dot \beta_2 &= -j\frac{\omega}{2} (\beta_1 - \beta_2) - j X I_t - V_o - \frac{1}{2} \kappa \beta_1 - \frac{1}{2} \kappa \beta_2 \notag \\
    \dot \xi &= -\kappa \xi + j\omega_\mathsf{nom} \xi + K_{iv} \frac{\beta_1 + \beta_2}{2} \notag \\
    m &= \matr K \Big[ I_b,\, V_o,\, I_t,\, \textstyle \frac{\beta_1 + \beta_2}{2},\,  \xi\Big]^\tran. \label{E:modified}
\end{align} 
Eq. (\ref{E:3}) is transformed into
\begin{align}
    L_c \dot I_b &= -\kappa L_c I_b - R_c I_b + V_o - V_b \notag \\
    C_f \dot V_o &= -\kappa C_f V_o + I_t - I_b \notag \\
    L_f \dot I_t &= -\kappa L_f I_t - R_f I_t + \frac{1}{2} V_{dc}  m - V_o. \label{E:rewritten_equations}
\end{align}

Define the state vector
\begin{align*}
    \vect x = \big[x_1,\ldots,\, x_6\big]^\tran &= \Big[L_c I_b,\, C_f V_o,\, L_f I_t,\, \beta_1,\, \beta_2,\, \xi\Big]^\tran.
\end{align*}
Define the Hamiltonian function as $H(\vect x) = \frac{1}{2} \| \vect x\|_{\matr Q}^2$
with 
\begin{equation*}
    \matr Q = \mathrm{diag}\big(L_c^{-1},\, C_f^{-1},\, L_f^{-1},\, 1,\, 1,\, 1\big).
\end{equation*}
The gradient is
\begin{align}
    &\nabla H(\vect x) = \matr Q \vect x = \Big[I_b,\, V_o,\, I_t,\, \beta_1,\, \beta_2,\, \xi \Big]^\tran. \label{E:grad}
\end{align}
Define the input and output as $\vect u_1 = V_b$ and $\vect y_1 = -I_b$. 
We can then compile (\ref{E:rewritten_equations}) and (\ref{E:modified}) into the pH system:
\begin{subequations} \label{E:pH_form} \begin{align}
    \dot{\vect x} &= \matr F \nabla H(\vect x) + j \frac{\omega}{2} \matr F_3 \vect x + \vect G \vect u_1 \\
    \vect y_1 &= \vect G^* \nabla H(\vect x), 
\end{align}\end{subequations}
where
\begin{align}
    &\matr F = \matr F_1 - \kappa \matr F_2, \matr F_1 = \matr F_0 + \matr B \matr K_1, \notag \\
    &\matr F_0 = \begin{bmatrix}
        -R_c &1 &0 &0 &0 &0 \\
        -1 &0 &1 &0 &0 &0 \\
        0 &-1 &- R_f &0 &0 &0 \\
        0 &-1 &-j X &0 &0 &0 \\
        0 &-1 &-j X &0 &0 &0 \\
        0 &0 &0 &\frac{K_{iv}}{2} &\frac{K_{iv}}{2} &j \omega_\mathsf{nom}
    \end{bmatrix}, \notag \\
    &\matr F_2 = \textstyle \mathrm{diag}\big(1,\, 1,\, 1,\, \big[\frac{3}{2},\, -\frac{1}{2};\frac{1}{2},\, \frac{1}{2}\big],\, 1 \big), \notag \\
    &\matr F_3 = \mathrm{diag}\big(0,\, 0,\, 0,\, \big[1,\, -1;-1,\, 1\big],\, 0\big),\, \notag \\
    &\matr G = {-}\vect e_1,\, \matr B = \frac{V_{dc}}{2} \vect e_3. \label{E:DER_matrices1}
\end{align}

\begin{rem} \label{rem_K}
The gain $\matr K_1$ in (\ref{E:DER_matrices1}) is equivalent to the gain $\matr K$ in (\ref{E:rewritten_equations}): The $4$-th and $5$-th columns of $\matr K_1$ are equal to $\frac{1}{2}$ times the $4$-th column of $\matr K$. The remaining columns of $\matr K_1$ are equal to those of $\matr K$ at appropriate positions.
\end{rem}

\section{Energy balance equations} \label{sec_eb}

The goal is to derive the energy balance for the shifted Hamiltonian defined with respect to the limit cycle $\eqm{\vect x}(\tau)$ that shrinks exponentially at the rate of $\kappa$. The phase $\tau$ is initially assumed to be a constant.

\subsection{Energy balance for each edge}

\textbf{Edges for DER generators.}
Analogous to the Lyapunov method in linear control,\footnote{The subscript $\texttt e$ is dropped for the DER subsystem.} we will derive the energy balance for the candidate storage function
$\hat H(\vect x) = \frac{1}{2} \| \vect x\|_{\hat{\matr Q}}^2$
where $\hat{\matr Q} = \matr P \matr Q = \matr Q \matr P^* \succ 0$ and
\begin{equation} \label{E:struct_P}
    \matr P = \mathrm{diag}(1,\, \matr P_{22})
\end{equation}
for $\matr P_{22} \in \mathbb{C}^{5\times 5}$.
The chosen structure of $\matr P$ in (\ref{E:struct_P}) preserves the output: $-I_b e^{j\theta} = \matr G^* \nabla H(\vect x) = \matr G^* \nabla \hat H(\vect x)$. Define the co-energy states $\vect s = \matr Q \vect x$ and $\eqm{\vect s}(\tau) = \matr Q \eqm{\vect x}(\tau)$. We make the following assumption to reduce the space of candidate storage functions to search.

\begin{assum} \label{assum_P}
Assume that the $4$-th row of $\hat{\matr Q} = \matr P \matr Q$ is equal to the $5$-th row except for the $2$-by-$2$ diagonal block, which is equal to $q_4 \matr I_2$ for $q_4 > 0$.
\end{assum}

Based on (\ref{E:s1-1}), the energy balance equation for the candidate storage function writes
\begin{align}
    &\dot{\hat{\mathcal H}}(\vect x, \eqm{\vect x}(\tau)) = \langle \nabla \hat H(\vect x) - \nabla \hat H( \eqm{\vect x}(\tau)), \dot{\vect x} - \dot{\eqm{\vect x}}(\tau) \rangle \notag \\
    &= \langle \nabla \hat H(\vect x) - \nabla \hat H(\eqm{\vect x}(\tau)), \matr F [\nabla H(\vect x) - \nabla H(\eqm{\vect x}(\tau)) ] \rangle \notag \\
    &\quad\, \textstyle + \langle \nabla \hat H(\vect x) - \nabla \hat H( \eqm{\vect x}(\tau)), \matr F_3 \big[j \frac{\omega}{2} \vect x - j \frac{\eqm\omega}{2} \eqm{\vect x}(\tau)\big] \rangle \notag \\
    &\quad\, + \langle \nabla \hat H(\vect x) - \nabla \hat H(\eqm{\vect x}(\tau)), \matr G \big[ \vect u - \eqm{\vect u}(\tau) \big] \rangle \notag \\
    &= \langle \vect s - \eqm{\vect s}(\tau), \matr P^* \matr F \big[\vect s - \eqm{\vect s}(\tau) \big] \rangle \notag \\
    &\quad\, \textstyle + \langle \vect x - \eqm{\vect x}(\tau), \matr Q \matr P^* \matr F_3 \big[j \frac{\omega}{2} \vect x - j \frac{\eqm\omega}{2} \eqm{\vect x}(\tau)\big] \rangle \notag \\
    &\quad\, + \langle \vect y_{1} - \eqm{\vect y}_{1}(\tau), \vect u_{1} - \eqm{\vect u}_{1}(\tau) \rangle. \label{E:s10-3}
\end{align}
The $2$-nd term in the RHS of (\ref{E:s10-3}) can be derived as
\begin{align}
    &\langle \vect x - \eqm{\vect x}(\tau), \textstyle \matr Q \matr P^* \matr F_3 \big[j\frac{\omega}{2} \vect x - j \frac{\eqm\omega}{2} \eqm{\vect x}(\tau) \big] \rangle \notag \\
    &= \langle \vect x - \eqm{\vect x}(\tau), \textstyle q_4 \matr F_3 \big[j\frac{\omega}{2} \vect x - j \frac{\eqm\omega}{2} \eqm{\vect x}(\tau) \big] \rangle \notag \\
    &= 2 q_4 \langle \varphi - \eqm\varphi(\tau), \textstyle j \omega \varphi - j \eqm\omega \eqm \varphi(\tau) \rangle. \label{E:s10-4}
\end{align}
Based on (\ref{E:s10-3}), (\ref{E:s10-4}), and the definition (\ref{E:widetilde}), the energy balance equation for $\widetilde{\mathcal H}(\vect x, \eqm{\vect x}(\tau))$ is obtained as
\begin{align}
    \dot{\hat{\mathcal H}}(\vect x, \eqm{\vect x}(\tau)) &= \|\vect s - \eqm{\vect s}(\tau) \|_{\mathrm{He}\{ \matr P^* \matr F\}}^2 \notag \\
    &\quad\, + 2 q_4 \langle \varphi - \eqm\varphi(\tau), \textstyle j \omega \varphi - j \eqm\omega \eqm \varphi(\tau) \rangle \notag \\
    &\quad\, + \langle \vect y_{1} - \eqm{\vect y}_{1}(\tau), \vect u_{1} - \eqm{\vect u}_{1}(\tau) \rangle. \label{E:sp_der}
\end{align}
It will be shown that the first term in (\ref{E:sp_der}) can be made negative definite with an appropriate choice of $\matr K_1$. 
For the second term in (\ref{E:sp_der}), we will defining a function $\hat\tau$ of the state of the overall microgrid and substituting $\tau = \hat\tau$ such that the second term in (\ref{E:sp_der}) cancels with $\frac{\partial}{\partial \tau} \hat{\mathcal H}(\vect x, \eqm{\vect x}(\hat\tau)) \frac{d}{dt} \hat\tau$. The function $\hat\tau$ is defined later in Subsection~\ref{sec_overall}. Here, we can gain more insight into the shifted Hamiltonian by noting the following decomposition:
\begin{equation} \label{E:widetilde}
    \hat{\mathcal H}(\vect x, \eqm{\vect x}(\tau)) = \widetilde{\mathcal H}(\vect x, \eqm{\vect x}(\tau)) + q_4 \|\varphi - \eqm\varphi(\tau)\|^2,
\end{equation}
where we have $\widetilde{\mathcal H}(\vect x, \eqm{\vect x}(\tau)) = \frac{1}{2}\|\vect x - \eqm{\vect x}(\tau)\|_{\widetilde{\matr Q}}^2$, and $\widetilde{\matr Q}$ is modified from $\hat{\matr Q}$ by replacing the diagonal block $q_4 \matr I_2$ with $q_4 \big[\frac{1}{2},\, \frac{1}{2};\frac{1}{2},\, \frac{1}{2}\big]$. Equivalently,
\begin{equation*}
    \widetilde{\matr Q} = \matr N \matr N^\dagger \hat{\matr Q} \matr N \matr N^\dagger
\end{equation*}
where
\begin{equation*}
    \matr N = \mathrm{diag}\big(1,\, 1,\, 1,\, \textstyle \big[1; 1 \big],\, 1\big).
\end{equation*}
Hence, $\widetilde{\matr Q} \succeq 0$ and $\widetilde{\mathcal H}(\vect x, \eqm{\vect x}(\tau)) \geq 0$. Clearly, $\widetilde H(\vect x, \eqm{\vect x}(\tau))$ is dependent on $\frac{1}{2}(\beta_1 + \beta_2) = \beta$ and is independent of $\frac{1}{2}(\beta_1 - \beta_2) = \varphi$.

\textbf{Edges for $R$--$L$ lines.} 
By (\ref{E:R-L}) and (\ref{E:energy_balance_pre}), the energy balance for every $R$--$L$ line edge is
\begin{align}
    \dot{\mathcal H}_\texttt{e}(\vect x_\texttt{e}, \eqm{\vect x}_\texttt{e}(\tau)) &= -(\kappa L_{\texttt e} + R_\texttt{e}) \|I_\texttt{e} - \eqm I_{\texttt e}(\tau) \|^2 \notag \\
    & + \langle \vect y_{\texttt{e},1} - \eqm{\vect y}_{\texttt{e},1}(\tau), \vect u_{\texttt{e},1} - \eqm{\vect u}_{\texttt{e},1}(\tau) \rangle. \label{E:eb_rl}
\end{align}

\textbf{Edges for shunt capacitors.}
By (\ref{E:shunt}) and (\ref{E:energy_balance_pre}), the energy balance for every shunt capacitor edge is
\begin{align}
    \dot{\mathcal H}_\texttt{e}(\vect x_\texttt{e}, \eqm{\vect x}_\texttt{e}(\tau)) &= -(\kappa C_{\texttt e} + G_{\texttt e}) \|V_{\texttt e} - \eqm{V}_{\texttt e}(\tau)\|^2 \notag \\
    &\hspace{-0.08in} + \langle V_\texttt{e} - \eqm{V}_\texttt{e}(\tau), \Upsilon_\texttt{e}'(t, V_\texttt{e}) - \Upsilon_\texttt{e}'(t, \eqm{V}_\texttt{e}(\tau)) \rangle \notag \\
    &\hspace{-0.08in} + \langle \vect y_{\texttt{e},1} - \eqm{\vect y}_{\texttt{e},1}(\tau), \vect u_{\texttt{e},1} - \eqm{\vect u}_{\texttt{e},1}(\tau) \rangle, \label{E:eb_cap}
\end{align}
where $\Upsilon_{\texttt e}'(t, V) = e^{-\kappa (t - t_0)} \Upsilon_{\texttt e}(e^{\kappa (t - t_0)} V)$.
For simplicity, we assume that all loads are globally shifted passive. On the other hand, if not all loads are globally passive~\cite{strehle2020passivity}, it is relatively straightforward to estimate the region of attraction by considering the shifted passivity boundaries of the loads.

\begin{assum} \label{assum_load}
Assume that the load $\Upsilon_\textup{\texttt{e}}(V_{\texttt e})$ is globally shifted passivity with respect to all $\eqm V_\textup{\texttt e} \in \mathbb{C}$.
\end{assum}

Then, it implies 
\begin{align*}
    &\langle V_\texttt{e} - \eqm{V}_\texttt{e}(\tau), \Upsilon_\texttt{e}'(t, V_\texttt{e}) - \Upsilon_\texttt{e}'(t, \eqm{V}_\texttt{e}(\tau)) \rangle \\
    &= \langle e^{-\kappa(t-t_0)} V_\texttt{e} - e^{-\kappa(t-t_0)}\eqm{V}_\texttt{e}(\tau),\, ... \notag \\
    &\hspace{0.25in} \Upsilon_\texttt{e}(e^{-\kappa(t-t_0)}V_\texttt{e}) - \Upsilon_\texttt{e}(e^{-\kappa(t-t_0)}\eqm{V}_\texttt{e}(\tau)) \rangle \leq 0.
\end{align*}

\subsection{Overall energy balance} \label{sec_overall}
Define the shifted Hamiltonian of the overall microgrid as
$\mathcal H = \sum_{\texttt e = 1}^{\texttt g} \hat{\mathcal H}_{\texttt e} + \sum_{\texttt e = \texttt g + 1}^{\texttt n} \mathcal H_{\texttt e}$.
Summing the energy balance equations for every edge, the input shifted power cancels as a consequence of the skew-symmetric interconnection constraints in (\ref{E:edge_constraints}), i.e.,
\begin{align*}
    &\langle \mathrm{col}\big(\vect y_{\texttt e, 1} - \eqm{\vect y}_{\texttt e, 1}(\tau)\big), \mathrm{col}\big(\vect u_{\texttt e,1} - \eqm{\vect u}_{\texttt e,1}(\tau)\big) \rangle \\
    &= \langle \mathrm{col}\big(\vect y_{\texttt e, 1} - \eqm{\vect y}_{\texttt e, 1}(\tau)\big), \matr W \big[\mathrm{col}\big(\vect y_{\texttt e, 1} - \eqm{\vect y}_{\texttt e, 1}(\tau)\big)\big] \rangle= 0.
\end{align*}
We then obtain the overall energy balance as
\begin{align}
    &\dot{\mathcal H}(\vect x, \eqm{\vect x}(\tau)) \leq \sum_{\texttt e=1}^{\texttt g} \|\vect s_{\texttt e} - \eqm{\vect s}_{\texttt e}(\tau) \|_{\mathrm{He}\{ \matr P_{\texttt e}^* \matr F_{\texttt e}\}}^2 \notag \\
    &\quad\, + \sum_{\texttt e=1}^{\texttt g} 2 q_{\texttt e,4} \langle \varphi_{\texttt e} - \eqm\varphi_{\texttt e}(\tau), j\omega_{\texttt e} \varphi_{\texttt e} - j \eqm\omega \eqm\varphi_{\texttt e}(\tau) \rangle \notag \\
    &\quad\, - \sum_{\texttt e = \textup{\texttt{g}} + 1}^{\textup{\texttt{g}} + \textup{\texttt{T}}} (\kappa L_{\texttt e} + R_\texttt{e}) \|I_\texttt{e} - \eqm I_{\texttt e}(\tau) \|^2 \notag \\
    &\quad\, - \sum_{\texttt e = \texttt g + \texttt T + 1}^{\texttt n} (\kappa C_{\texttt e} + G_{\texttt e}) \|V_{\texttt e} - \eqm V_{\texttt e}(\tau)\|^2. \label{E:overall_eb}
\end{align}
As mentioned above, let us define $\hat\tau$ as a solution of the following time-varying differential equation:
\begin{align}
    0 &= \frac{\partial}{\partial \tau} \mathcal H(\vect x, \eqm{\vect x}(\tau))\, \dot{\tau}\, ... \notag \\
    &+ \sum_{\texttt e=1}^{\texttt g} 2 q_{\texttt e,4} \langle \varphi_{\texttt e} - \eqm\varphi_{\texttt e}(\tau), j\omega_{\texttt e} \varphi_{\texttt e} - j \eqm\omega \eqm\varphi_{\texttt e}(\tau) \rangle, \label{E:s11-1}
\end{align}
which can be expressed as
\begin{align}
    0 &= \Bigg[\sum_{\texttt e=1}^{\texttt g} \langle \vect x_{\texttt e}, j \widetilde{\matr Q}_{\texttt e} \eqm{\vect x}_{\texttt e}(\tau) \rangle + \sum_{\texttt e=\texttt g+1}^{\texttt g+\texttt T} \langle I_{\texttt e}, j L_{\texttt e} \eqm I_{\texttt e}(\tau) \rangle\, ... \notag \\
    &+ \sum_{\texttt g+\texttt T+1}^{\texttt n} \langle V_{\texttt e}, j G_{\texttt e} \eqm V_{\texttt e}(\tau) \rangle \Bigg] \dot\tau + \frac{d}{dt} \sum_{\texttt e=1}^{\texttt n} q_{\texttt e,4} \|\varphi_{\texttt e} - \eqm\varphi_{\texttt e}(\tau)\|^2, \label{E:s11-2}
\end{align}
where the last derivative is with respect to the original system (without the shrinking effect). Note that (\ref{E:s11-1}) is solvable in the set $E = \{ \vect x\mid \frac{\partial}{\partial \tau} \mathcal H(\vect x, \eqm{\vect x}(\tau)) \neq 0 \}$, which is almost everywhere in $\mathbb{C}^n$ ($n = 8\texttt g + 2\ell + 2\texttt T$). Then, we make the following simplifying assumption.

\begin{assum} \label{assum_sol}
Assume every solution of the microgrid system belongs to the set $E$ for almost all $t\in \mathbb{R}$.
\end{assum}

Assumption~\ref{assum_sol} is justified by noting that the Hamiltonian can be slightly perturbed independently of the behavior of the system to obtain a different set $E$.

Substituting (\ref{E:s11-1}) into (\ref{E:overall_eb}), we obtain
\begin{align}
    &\dot{\mathcal H}(\vect x, \eqm{\vect x}(\hat\tau)) \leq \sum_{\texttt e=1}^{\texttt g} \|\vect s_{\texttt e} - \eqm{\vect s}_{\texttt e}(\hat\tau) \|_{\mathrm{He}\{ \matr P_{\texttt e}^* \matr F_{\texttt e}\}}^2 \notag \\
    &\quad\, - \sum_{\texttt e = \textup{\texttt{g}} + 1}^{\textup{\texttt{g}} + \textup{\texttt{T}}} (\kappa L_{\texttt e} + R_\texttt{e}) \|I_\texttt{e} - \eqm I_{\texttt e}(\hat\tau) \|^2 \notag \\
    &\quad\, - \sum_{\texttt e = \texttt g + \texttt T + 1}^{\texttt n} (\kappa C_{\texttt e} + G_{\texttt e}) \|V_{\texttt e} - \eqm V_{\texttt e}(\hat\tau)\|^2. \label{E:overall_eb2}
\end{align}

\subsection{Main result} \label{sec_main}

The main stability result is stated as follows. Define the notation $\check{\matr F}_{\texttt e,1} = \matr N^\dagger \matr F_{\texttt e,1} \matr N$ and $\check{\matr P}_{\texttt e} = \matr N^* \matr P_{\texttt e} \matr N$.

\begin{prop} \label{prop_main}
Assume that $\mathrm{He}\{\check{\matr P}_\textup{\texttt e}^* \check{\matr F}_{\textup{\texttt e}, 1}\} \prec 0$ for $\textup{\texttt e} = 1,\ldots,\, \textup{\texttt g}$. 
Assume that the circular limit cycle $\eqm{\vect x}(\tau)$ from Assumption~\ref{assum_lim} is the only circular limit cycle of the system.
Then, the orbit of $\eqm{\vect x}(\tau)$ is globally attractive.
\end{prop}

\begin{pf}
From (\ref{E:overall_eb2}), we have that $\dot{\mathcal H}(\vect x, \eqm{\vect x}(\hat\tau)) \leq 0$ if, for $\texttt e = 1,\ldots,\, \texttt g$, it holds that
\begin{equation} \label{E:question}
    \mathrm{He}\{\matr P_{\texttt e}^* \matr F_{\texttt e} \} \prec 0,
\end{equation}
which can be expanded into
\begin{equation} \label{E:ineq2}
    \mathrm{He}\{\matr P_{\texttt e}^* \matr F_{\texttt e, 1}\} - \kappa \mathrm{He}\{\matr P_{\texttt e}^* \matr F_{\texttt e, 2}\} \prec 0
\end{equation}
Note that the row and column spaces of the matrix $\mathrm{He}\{\matr P_{\texttt e}^* \matr F_{\texttt e, 1}\}$ are equal to the column space of $\matr N$. By the assumption, there exists an $a > 0$ such that
\begin{equation} \label{E:ineq3}
    \mathrm{He}\{\check{\matr P}_{\texttt e}^* \check{\matr F}_{\texttt e, 1}\} + a (\matr N^\dagger)^* \matr N^\dagger \prec 0.
\end{equation}
Now, to prove (\ref{E:ineq2}), it suffices to prove that 
\begin{equation} \label{E:ineq4}
    -a (\matr N^\dagger)^* \matr N^\dagger - \kappa \mathrm{He}\{\matr P_{\texttt e}^* \matr F_{\texttt e,2}\} \prec 0.
\end{equation}
It is not difficult to verify, from the definition of $\matr F_{\texttt e, 2}$ in (\ref{E:DER_matrices1}) and Assumption~\ref{assum_P}, that all elements on the main diagonal of $\mathrm{He}\{\hat{\matr Q}_{\texttt e} \matr F_{\texttt e,2}\} = \mathrm{He}\{\matr Q_{\texttt e} \matr P_{\texttt e}^* \matr F_{\texttt e,2}\}$ are positive, and the same holds for $\mathrm{He}\{\matr P_{\texttt e}^* \matr F_{\texttt e, 2}\}$. Now, consider the following decomposition:
\begin{equation*}
    2 \mathrm{He}\{\matr P_{\texttt e}^* \matr F_{\texttt e, 2}\} = \matr Q_1 + \matr Q_2,
\end{equation*}
where $\matr Q_1$ keeps the $4$-th and $5$-th rows and columns of $2 \mathrm{He}\{\matr P_{\texttt e}^* \matr F_{\texttt e, 2}\}$, and $\matr Q_2$ keeps the rest. Denote by $\matr N_1$ the identity matrix with the $4$-th and $5$-th diagonal element replaced by zero. We obtain the following sufficient condition for (\ref{E:ineq4}):
\begin{align}
    \begin{cases}
        -a \matr N_1 - \kappa \matr Q_1 \prec 0, \\
        -a \matr N_1 - \kappa \matr Q_2 \preceq 0.
    \end{cases} \label{E:twin}
\end{align}
We claim that condition (\ref{E:twin}) holds for small enough $\kappa$.
For the first inequality, we can take the Schur complement of the $2$-by-$2$ diagonal block between the $4$-th and $5$-th row of the LHS to see that it holds for small enough $\kappa$. By construction, the second inequality holds for small enough $\kappa$. Hence, (\ref{E:twin}) holds for small enough $\kappa$, which implies (\ref{E:question}).

Recall from Section~\ref{sec_PH} that the solutions of the system are artificially shrunk by $e^{-\kappa(t - t_0)}$. We can remove this artificial shrinking from the energy balance equation to obtain
$\dot{\mathcal H}(\vect x, \eqm{\vect x}(\hat\tau)) \leq 2\kappa \mathcal H(\vect x, \eqm{\vect x}(\hat\tau))$,
or equivalently,
\begin{equation} \label{E:ineq7}
    \frac{d}{dt} \log(\mathcal H(\vect x, \eqm{\vect x}(\hat\tau))) \leq 2\kappa.
\end{equation}
Notice that, if we fix the scale of $\varphi$, from (\ref{E:widetilde}) and (\ref{E:shrink}), we can express $\mathcal H(\vect x, \eqm{\vect x}(\tau))$ as
\begin{equation} \label{E:ineq8}
    \mathcal H(\vect x, \eqm{\vect x}(\tau)) = \widetilde{\mathcal H}(\vect x, \eqm{\vect x}(\tau)) + \sum_{\texttt e=1}^{\texttt g} \kappa^{-2} q_4 \|\varphi - \eqm\varphi(\tau)\|^2,
\end{equation}
where $\widetilde{\mathcal H}(\vect x, \eqm{\vect x}(\tau))$ is a quadratic function not dependent on $\kappa$. Notice also that, by the definition of $\hat\tau$, the derivative of the second term in (\ref{E:ineq8}) is canceled. Hence (\ref{E:ineq7}) can be expressed as
\begin{equation} \label{E:ineq9}
    \frac{d}{dt} \log(\widetilde{\mathcal H}(\vect x, \eqm{\vect x}(\hat\tau))) \leq 2\kappa,
\end{equation}
By taking $\kappa \to 0$, $\hat\tau$ converges to a certain function according to (\ref{E:s11-2}), and we obtain that $\dot{\widetilde{\mathcal H}}(\vect x, \eqm{\vect x}(\hat\tau)) \leq 0$ in the limit.
Then, applying the monotone convergence theorem, we obtain that $\widetilde{\mathcal H}(\vect x, \eqm{\vect x}(\hat\tau))$ is convergent as $t\to \infty$.

Note that $\widetilde{\mathcal H}(\vect x, \eqm{\vect x}(\tau))$ is a quadratic function in the form of $\|\vect x - \eqm{\vect x}(\hat\tau)\|_{\matr Q}^2$. From the above, the limiting solution satisfies
\begin{equation} \label{E:equi_d}
    \|\vect x - \eqm{\vect x}(\hat\tau)\|_{\matr Q} = c.
\end{equation}
for some constant $c$. 
Denote by $n = 8\texttt g + 2\ell + 2\texttt T$ the dimension of $\vect x$, i.e., $\vect x \in \mathbb{C}^n$. Note that the inequality (\ref{E:ineq9}) only depends on the linear part of the system, i.e., decoupled from $\varphi_{\texttt e}$, and the linear part of the system is Hurwitz from (\ref{E:overall_eb2}) and the condition in the proposition statement, 
Hence, without violating Assumption~\ref{assum_P},\footnote{Assumption~\ref{assum_P} implies that $\widetilde{\mathcal H}(\vect x, \eqm{\vect x}(\hat{\tau}))$ and $\dot{\widetilde{\mathcal H}}(\vect x, \eqm{\vect x}(\hat\tau))$ are independent of $\varphi_{\texttt e}$ and $\varphi_{\texttt e}(\hat\tau)$.} we can perturb $\matr Q$ to reach the same inequality (\ref{E:ineq9}). Let us consider the norm $\|\vect y\|_{ij} = \sqrt{\langle \vect y, \mathrm{He}\{\matr Q + \epsilon \hat{\matr N} \vect e_i \vect e_j^* \hat{\matr N}^* \}\vect y \rangle}$ for $0 < \epsilon \ll 1$ and $1\leq i\leq j\leq n - \texttt g$, where $\hat{\matr N} = \mathrm{diag}\big(\matr I_{\texttt g} \otimes \matr N,\, \matr I_{\texttt g + \ell + \texttt T}\big)$.
We then obtain\footnote{The convergence of $\varphi_{\texttt e} - \eqm\varphi_{\texttt e}(\hat\tau)$ at the limiting solution can be proved by the invariance principle.}
\begin{equation}\label{E:eq3}
    \langle x_i - \eqm x_i(\hat\tau), x_j - \eqm x_j(\hat\tau)\rangle = c_{ij}
\end{equation}
for some constant $c_{ij}$. Note that (\ref{E:eq3}) entails $2n-1$ real constraints on the limiting solution, which include 
\begin{itemize}
    \item constant $\|x_i(t)- \eqm x_i(\hat\tau)\|$,
    \item constant $\arg(x_i(t) - \eqm x_i(\hat\tau)) - \arg(x_i(t) - \eqm x_j(\hat\tau))$.
\end{itemize}
Then, the limiting solution can only evolve in the direction of the angle symmetry, i.e., it has a circular orbit. By the assumption on the uniqueness of the circular orbit of $\eqm{\vect x}(\tau)$, it is the only positive limit set, which completes the proof.
\hfill $\square$
\end{pf}

Three comments are in order. \textcolor{myblue}{Let us consider the scenario where $\kappa$ is small. On one hand, the shifted Hamiltonian from (\ref{E:widetilde}) is dominated by $\sum_{\texttt e=1}^{\texttt g} q_{\texttt e,4} \|\varphi_{\texttt e} - \eqm\varphi_{\texttt e}(\tau)\|^2$ since $|\varphi_{\texttt e}| \sim 1/\kappa$. On the other hand, the time-varying phase $\hat\tau$ in (\ref{E:s11-2}) is defined to cancel the time derivative of $\sum_{\texttt e=1}^{\texttt g} q_{\texttt e,4} \|\varphi_{\texttt e} - \eqm\varphi_{\texttt e}(\tau)\|^2$ by ``averaging'' the DER frequencies. Here, the ``average'' is defined in terms of $p_{\texttt e,4}$, i.e., the diagonal elements of the Hamiltonian associated with the state variable $\beta_{\texttt e}$, rather than the inertia constants $J_{\texttt e}$~\cite{liu2016power}. This ``averaging'' essentially isolates the phase synchronization process from the energy dissipation process. The latter allows us to prove the attractivity of the periodic orbit.}

\textcolor{myblue}{To remove the effect of shrinking, the shifted passivity condition in Proposition~\ref{prop_main} only implies that the shifted Hamiltonian is non-increasing. This is in contrast with the equilibrium point case, where the dissipation of energy is directly correlated to the quadratic dissipation matrix. The explanation is as follows: The system has angle symmetry, which means that the circular limit cycle cannot be attractive in phase. Since two limit cycles with different phases maintain a constant distance from each other, the maximal condition to impose on the energy function is for it to be non-increasing. Hence, stability is not proved in terms of the usual Lyapunov stability of the periodic orbit as a compact set~\cite{yi2020orbital,chiang2011direct}.}

The proposed control ensures global attractivity of the periodic orbit, which implies frequency synchronization and voltage stability, regardless of the inertia constant $J_{\texttt e}$ of the grid-forming DERs. This is a significant departure from the operational philosophy of the traditional power system, which relies on inertia to guarantee system stability. Since inertia does not affect stability with the proposed control, smaller inertia may lead to shorter settling times for power and frequency, with respect to the time constants of the distribution lines. 
It would be a potential future direction to study the optimal combination of $J_{\texttt e}$ and $D_{\texttt e}$ for more effective power sharing and smaller frequency deviation, based on the proposed control scheme that separates their choice from the stability of the system.

\section{Design of the DER control gain} \label{sec_control_param}

From Proposition~\ref{prop_main}, the stability condition writes\footnote{The subscript $\texttt e$ is omitted in this section.}
\begin{subequations} \begin{align}
    &\mathrm{He}\{\check{\matr P}^* \check{\matr F}_1\} \prec 0, \label{E:s14-3} \\
    &\check{\matr P} \check{\matr Q} = \check{\matr Q}^* \check{\matr P}^* \succ 0. \label{E:s14-4}
\end{align} \end{subequations} 
where $\check{\matr Q} = \matr N^\dagger \matr Q \matr N$.
Define:
\begin{align*}
    &\check{\matr K}_1 = \matr K_1 \matr N = \matr K,\, \check{\matr F}_0 = \matr N^\dagger \matr F_0 \matr N,\, \check{\matr B} = \matr N^\dagger \matr B,
\end{align*}
and $\check{\matr Q}_{22}$ as the $4$-by-$4$ lower-right submatrix of $\check{\matr Q}$.
Left and right multiplying (\ref{E:s14-3}) by $(\check{\matr P}^{-1})^*$ and $\check{\matr P}^{-1}$, we obtain the equivalent condition
\begin{align}
    &\mathrm{He}\{\check{\matr F}_1 \check{\matr P}^{-1}\} = \mathrm{He}\{\check{\matr F}_{0} \check{\matr P}^{-1} + \check{\matr B} \matr K \check{\matr P}^{-1} \} \prec 0 \label{E:s14-2}
\end{align}
The control design problem can then be formulated~\cite{riverso2014plug} as the following linear matrix inequalities by choosing the unknowns as $\matr L = \matr K \matr P^{-1},\, \matr X = \matr X^* = (\matr P \matr Q)^{-1}$.

\begin{prob}[Design of the DER control gain] \label{prob_design}
\begin{align}
    &\underset{\matr L \in \mathbb{C}^{1\times 5}, \matr X_{22} = \matr X^*_{22} \in \mathbb{C}^{4\times 4} \succ 0}{\min}  c_1 \alpha + c_2 \zeta + c_3 \gamma\;\;\textup{s.t.} \notag \\
    &\begin{bmatrix}
        \mathrm{He}\big(\check{\matr F}_{0} \check{\matr P}^{-1} + \check{\matr B} \matr L\big) &(\check{\matr P}^{-1})^* \\
        \check{\matr P}^{-1} &-\gamma \matr I_{5}
    \end{bmatrix} \preceq 0, \label{E:prob_1} \\
    &\begin{bmatrix}
    \alpha \matr I_{5} &\mathbf{L}^* \\
    \mathbf{L} &1
    \end{bmatrix} \succeq 0,\, \begin{bmatrix}
    \matr Z_{22} &\mathbf{I}_4 \\
    \mathbf{I}_4 &\zeta\mathbf{I}_4
    \end{bmatrix} \succeq 0, \label{E:prob_2} \\
    &\alpha, \zeta, \gamma \geq 0,\, \check{\matr P}^{-1} = \mathrm{diag}(1,\, \check{\matr P}_{22}^{-1}),\, \check{\matr P}_{22}^{-1} = \check{\matr Q}_{22} \matr X_{22}. \notag
\end{align}
The resulting control gain is recovered by $\matr K = \matr L \check{\matr P}$.
\end{prob}

Regularization introduced in Problem~\ref{prob_design} are explained as follows:
\begin{enumerate}
\item[1)] Using Schur complement, we can rewrite (\ref{E:prob_1}) as
\begin{equation*}
    \check{\matr F}_0 \check{\matr P}^{-1} \preceq - \gamma^{-1} (\check{\matr P}^{-1})^* \check{\matr P}^{-1},
\end{equation*}
which is equivalent to 
$\check{\matr P}^* \check{\matr F}_1 \preceq -\gamma^{-1} \matr I_5$.
Hence by (\ref{E:overall_eb}), $\gamma^{-1}$ is a lower bound for the dissipation level of the DER.
\item[2)] The regularization in (\ref{E:prob_2}) can be rewritten as the following bounds,
\begin{align*}
    \matr L^* \matr L \preceq \alpha \matr I_{5},\, \matr Z_{22} \preceq \zeta \matr I_4.
\end{align*}
\end{enumerate}

\begin{rem}
The same idea of searching for another candidate quadratic Hamiltonian in control gain design can be found in~\cite{johnsen20082}. However, in~\cite{johnsen20082} only three simple matrix types are considered as candidates for $\check{\matr P}$ whereas the precise condition is $\check{\matr P} \check{\matr Q} = (\check{\matr P} \check{\matr Q})^* \succ 0$.
\end{rem}
\section{Numerical example} \label{sec_numerical}

The goal is to test the feasibility of the control gain design in Problem~\ref{prob_design}, to simulate the transient dynamics of a microgrid test system under perturbations that include load changes and disconnection/reconnection between different parts of the microgrid and the main grid, and to compare its transient responses with the standard droop control with the cascaded double loop PI control. The latter is designed in the frequency domain with the time scale separation assumption as specified in~\cite{pogaku2007modeling}.

\subsection{Verifying the feasibility of the control design} \label{sec_verification}

The parameters for Problem~\ref{prob_design} in this test are provided in Table~\ref{table_parameters}. All quantities are per unit except for the frequencies, which are in Hz. The $RLC$ filter parameters are chosen as in~\cite{pogaku2007modeling} to attenuate the switching frequency of the inverter. The steady-state droop characteristic of the control is determined by $D$ for the frequency droop and $X$ for the voltage droop, which are respectively chosen for the microgrid system to have $0.5\%$ nominal frequency droop at $0.1$ MW active power output and $5\%$ voltage droop at $0.075$ MVA reactive power output. The inertial $J$ is chosen so that the swing equation has a time constant of $0.1$~s. The matrices in (\ref{E:prob_1}) are obtained from (\ref{E:DER_matrices1}) with $\omega_{\mathsf{nom}} = 60$ Hz. The parameter $K_{iv}$ is chosen as $2\times 10^4$ because in control gain validation, we find that larger $K_{iv}$ results in a $\matr K$ that provides a faster convergence rate of the system under load changes. This value for $K_{iv}$ gives sufficiently fast convergence without causing $\matr K$ to become too large.

Problem~\ref{prob_design} is solved with MOSEK in YALMIP, which is feasible without numerical issues. In the solution obtained in this test, the eigenvalues of $\check{\matr P}_{22} \check{\matr Q}_{22}$ are between $0.0081$ and $3.7817\times 10^5$, and the eigenvalues of $\mathrm{He}\{\check{\matr P}^*\check{\matr F}_1\}$ are between $-715.08$ and $-0.035$.

\setlength{\arraycolsep}{1.35pt} 
\begin{table}[!t]
\scriptsize
\renewcommand{\arraystretch}{1.2}
\caption{Parameters for the numerical example}
\label{table_parameters}
\setlength\tabcolsep{2pt}
\centering
\resizebox{\columnwidth}{!}{
\begin{tabular}{|c|c|}
\hline
\bf Description &\bf Parameter \\
\hline
Per-unit base &$V_{\mathrm Bl\text{-}l}=400$ V, $P_{\mathrm B} = 1$ MW, $\omega_{\mathsf{nom}} = 60$ Hz \\
\hline
RLC filter &$R_f = 0.1$ $\Omega$, $L_f = 1.35$ mH, $C_f = 50$ $\mu$F \\
\hline
Line coupling &$R_c = 0.14$ $\Omega$, $L_c = \textcolor{myblue}{1.6\times 10^{-7}}$ mH \\
\hline
\multicolumn{2}{|c|}{\bf Proposed reference-frame based control} \\
\hline
Swing equation &$D = 0.05$ pu, $J = 2$ pu \\
\hline
Misc. &$X = 0.4$ pu, $V_n = 1$ pu, $V_{dc} = 2$ pu, \\
\hline
Hyperparam. & $K_{iv} = 2\times 10^4,\, c_1 = 1,\, c_2 = 10^2,\, c_3 = 10^5$ \\
\hline
\multicolumn{2}{|c|}{} \\[-1em]
\multicolumn{2}{|c|}{$\matr K = \Big[
    117 - j0.5,\, -129,\, -115 - j0.3,\, 1293 - j4471,\, \textcolor{myblue}{-49.9 - j11.6} \Big] $} \\[4pt]
\hline
\multicolumn{2}{|c|}{{\bf Baseline droop control} (designed with $P_{\mathrm B} = 0.1$ MW)} \\
\hline
Droop gains &$m_p = 0.005$ [60Hz], $n_q = 0.0667$ \\
\hline
Input filter cutoff & $\omega_c = 6$ Hz \\
\hline
Double loop &$K_{pv} = 0.1833$, $K_{iv} = 230.94$, \\
PI gains &$K_{pc} = 7.59$, $K_{ic} = 4.48\times 10^4$, $K_f = 0.75$ \\
\hline
\multicolumn{2}{|c|}{\bf Microgrid test system} \\
\hline
ZIP load &$Z_{ld} = 65.29 + j48.97$ pu, $I_{ld} = 0$ pu, \\
&$S_{ld} = (1 + j0.75) \times 10^{-3}$ pu \\
\hline
Shunt capacitor &$G_{sh} = 0$ S, $C_{sh} = 16$ $\mu$F  \\
\hline
\end{tabular}
}
\end{table}

\ctikzsubcircuitdef{DGU}{in 1, gnd1}{
    coordinate(#1-gnd1) node[tlground]{} 
    to[vsourcesin] ++(0,1)
    coordinate(#1-in 1) node[circ]{}
}
\ctikzsubcircuitdef{load}{in 1, gnd1, gnd2}{
    coordinate(#1-gnd1) node[tlground]{}
    to[vR,mirror] ++(0,1)
    coordinate(#1-in 1) node[circ]{} -- ++(0.4,-0.2) to[C] ++(0,-0.8)
    node[tlground]{} coordinate(#1-gnd2)
}

\ctikzsubcircuitactivate{DGU}
\ctikzsubcircuitactivate{load}

\begin{figure}
\centering
\scalebox{0.85}{
\begin{circuitikz}[european]
    \ctikzset{bipoles/length=0.5cm}
    \coordinate[label=right:\scriptsize Main grid] (B) at (5.8,-0.58);

    \draw (0,0) \load{nine}{};
    \draw (nine-in 1) node[above]{\scriptsize $9$} to[american inductor] ++(1.2,0)
        node[above]{\scriptsize $5$} \DGU{five}{in 1};
    \draw (five-in 1) to[american inductor] ++(1.2,0)
        node[above]{\scriptsize $10$} \load{ten}{in 1};
    \draw (ten-in 1) to[american inductor] ++(1.2,0)
        node[above]{\scriptsize $6$} \DGU{six}{in 1};
    \draw (six-in 1) to[normal closed switch]
        ++(1.2,0) to[american inductor] ++(0,-1.6) node[circ]{} to[normal open switch] ++(1,0);
    \draw (0,-0.6) node[above]{\scriptsize $8$} \DGU{eight}{in 1};
    \draw (eight-in 1) to[american inductor] ++(1.2,0)
        node[above]{\scriptsize $12$} \load{twelve}{in 1};
    \draw (twelve-in 1) to[american inductor] ++(1.2,0)
        node[above]{\scriptsize $7$} \DGU{seven}{in 1};
    \draw (seven-in 1) to[american inductor] ++(1.2,0)
        node[above]{\scriptsize $11$} \load{eleven}{in 1};
    \draw (eleven-in 1) to 
        ++(1.2,0);
    \coordinate[label=above:\scriptsize CB1] (B) at (4.2,1.1);
    \coordinate[label=above:\scriptsize CB2] (B) at (5.3,-0.5);
\end{circuitikz}}
    \caption{Topology of the microgrid test system (ZIP loads are presented by variable resistors)}
    \label{fig_test_case}
\end{figure}

\subsection{Comparing transient responses with the standard droop control}

In Fig.~\ref{fig_test_case}, the topology of the microgrid test system is shown. The single-phase electromagnetic model of the system with line dynamics is defined and numerically solved in MATLAB. 
At the initial condition, every bus injects or consumes the same $(1 + j 0.75) \times 10^{\unaryminus 2}$~pu complex power before adjustment for the $R$--$L$ line and shunt capacitor losses. The ZIP loads are configured to have $10\%$ constant power component at the initial condition so that they remain shifted passive if the voltage of the shunt capacitor does not drop below $0.13$ pu.

\begin{figure}[t]
\includegraphics[width=3.22in,center,margin=-0.01in 0.0in 0in 0in]{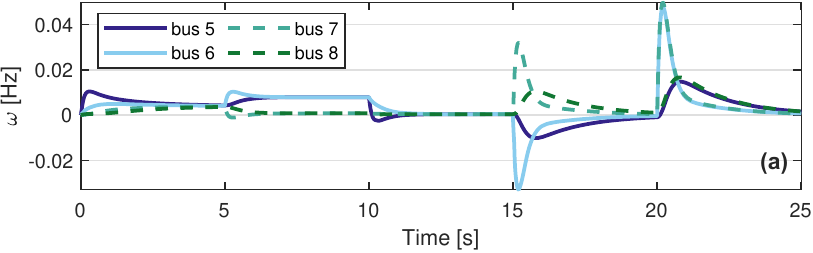}
\hfill
\includegraphics[width=3.26in,center,margin=-0.03in 0.0in 0in 0.0in]{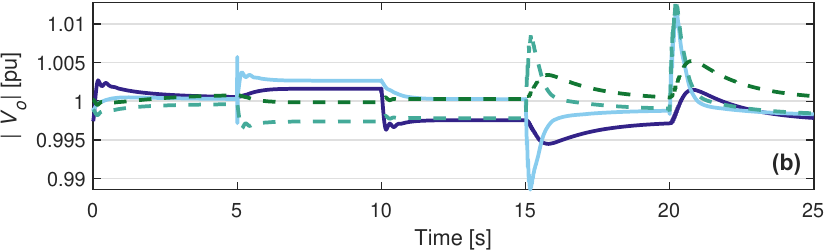}
\hfill
\includegraphics[width=3.24in,center,margin=-0.02in 0.0in 0in 0.0in]{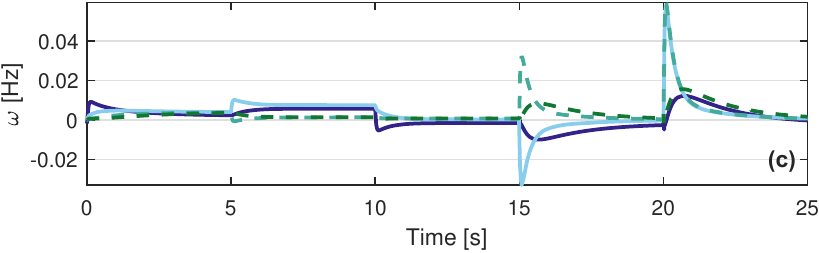}
\hfill
\includegraphics[width=3.24in,center,margin=-0.03in 0.0in 0in 0.0in]{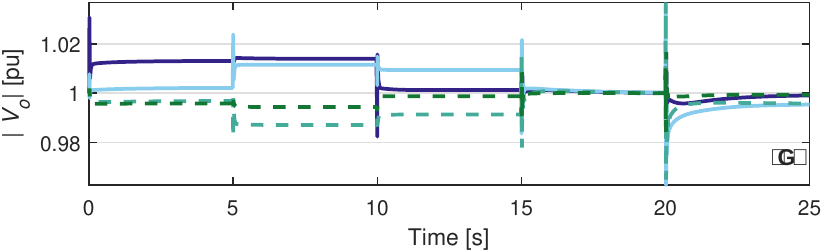}
\caption{Comparing the transient responses of the proposed control {\bf (a)--(b)} and double-loop PI control {\bf(c)--(d)} under the following disturbances: from $0.01$ to $10$ s, the conductance and susceptance of the Z load at Bus 9 are reduced by $40\%$ and $50\%$ respectively, the conductance of the Z load at Bus 10 is reduced by $20\%$, and the real power of the CPL at Bus 10 is increased by $40\%$. From $5$ to $15$ s, CB1 is opened. At $20$ s, CB2 to the infinite bus $V_{\inf} = 0.95 \angle 0.6$~pu is closed}
\label{fig_simulation}
\end{figure}

\textbf{Test 1.} For the first test, line resistance and inductance are set to $0.2$ $\Omega$ and $4$ mH.
The transient response of the proposed control is shown in Fig.~\ref{fig_simulation}a and \ref{fig_simulation}b under discrete disturbances which include load changes at $0.01$ and $10$~s, disconnection and reconnection of the microgrid topology at $5$ and $15$ s, and reconnection to a stiff main grid (an infinite bus) at $20$ s. The latter two are simulated as sudden actions of the circuit breakers without pre-synchronization to induce a large transient response. 
The frequency and voltage responses of the proposed controller in Fig.~\ref{fig_simulation}a and \ref{fig_simulation}b are comparable to the standard droop control in Fig.~\ref{fig_simulation}c and Fig.~\ref{fig_simulation}d. We can see that the frequency and voltage responses of the proposed controller are smooth and surprisingly similar in shape. For comparison, the frequency response of the standard droop control is smooth and makes sense from the first-order dynamics of the swing equation, but the voltage response has large overshoots, which indicates that the voltage is regulated on a smaller time scale.

\begin{figure}[t]
\includegraphics[width=3.18in,center,margin=0.005in 0.0in 0in 0in]{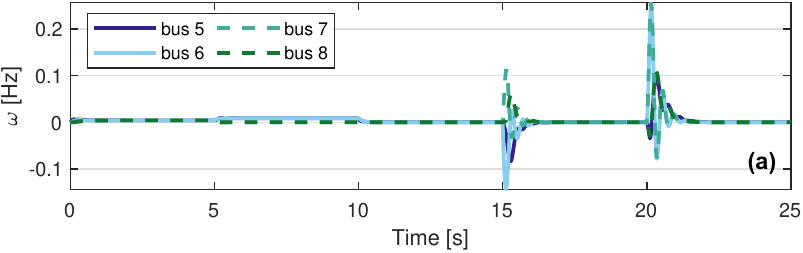}
\hfill
\includegraphics[width=3.21in,center,margin=-0.01in 0.0in 0in 0.0in]{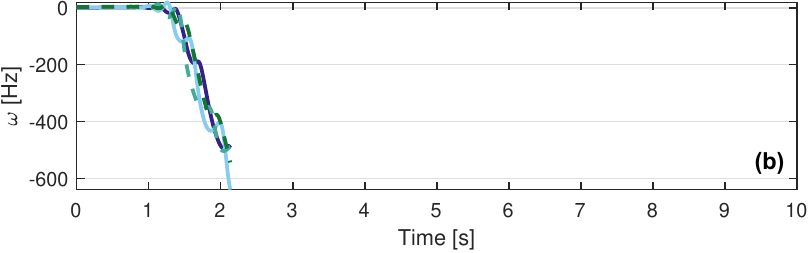}
\caption{Comparing the transient responses of the proposed control in \textbf{(a)} and baseline control in \textbf{(b)} with the time constant of the network $R$--$L$ lines reduced from $20$~ms to $1$~ms.}
\label{fig_frequency}
\end{figure}

\begin{figure}[t]
\includegraphics[width=3.28in,center,margin=-0.03in 0.0in 0in 0in]{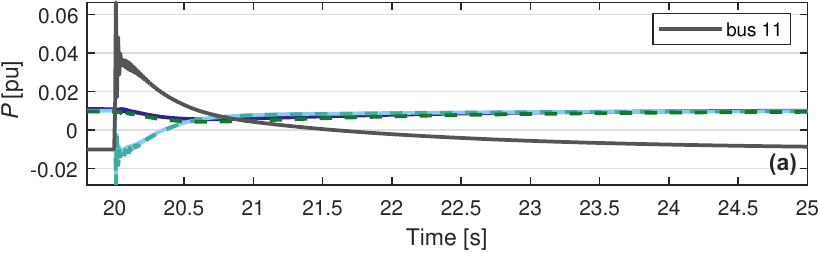}
\hfill
\includegraphics[width=3.28in,center,margin=-0.03in 0.0in 0in 0.0in]{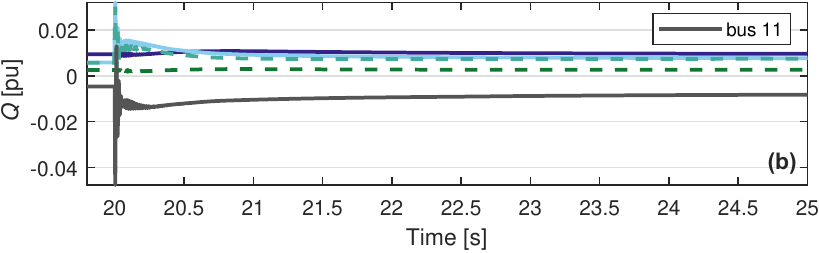}
\hfill
\includegraphics[width=3.23in,center,margin=0in 0.0in 0in 0.0in]{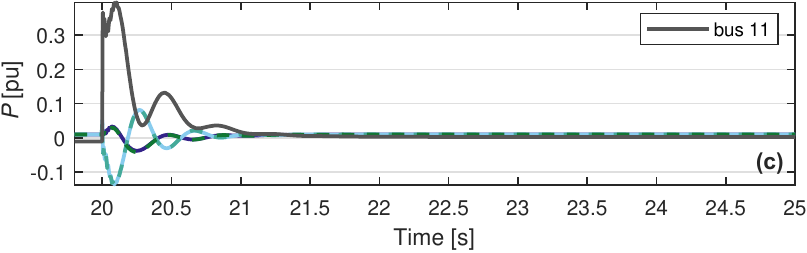}
\hfill
\includegraphics[width=3.23in,center,margin=0in 0.0in 0in 0.0in]{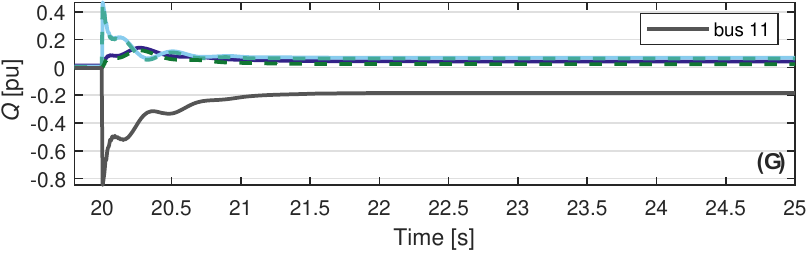}
\caption{Power sharing characteristic with the proposed control under reconnection to the stiff main grid without pre-synchronization, {\bf (a)--\bf (b)}: time constant of $R$--$L$ lines is $20$~ms, {\bf (c)--(d)}: time constant of $R$--$L$ lines is $1$~ms}
\label{fig_power_sharing}
\end{figure}

\textbf{Test 2.} The difference between the two types of control is more obvious when we change the line parameters to $0.1$ $\Omega$ and $0.1$ mH. These line parameters better represent a low-voltage microgrid where the distribution lines are mostly resistive. It corresponds to a decrease in the time constant of the $R$--$L$ line dynamics from $20$ ms to $1$ ms ($1$ kHz). The latter is on the same order of magnitude as the voltage and current loops of the standard control, which are, respectively, $0.4$ and $1.6$ kHz. Consistent with the analysis in~\cite{subotic2020lyapunov}, we can see in Fig.~\ref{fig_frequency} that without changing the control gains of the standard droop control, the trajectory fails to converge after the first disturbance at $0.1$~s. In comparison, the proposed controller maintains stability as the time constant of the $R$--$L$ lines is reduced $20$ times.

The power sharing characteristic of the proposed controller under the large disturbance of the reconnection event at $20$ s in the two tests is shown in Fig.~\ref{fig_power_sharing}a--d. In all cases, we can see that the large power disturbances at Bus~11 are shared during the transient so that the impact on each DER is lessened. In particular, in Fig.~\ref{fig_power_sharing}, a surge of real power is injected into Bus~11 at $20$ s, but the changes in DER power outputs are much milder. 
Comparing the power response with the two sets of line parameters, we can see that in Fig.~\ref{fig_power_sharing}a and \ref{fig_power_sharing}b, the settling times of the real and reactive power are indeed much longer than in Figs.~\ref{fig_power_sharing}c and \ref{fig_power_sharing}d, which agrees with the estimate that the response of the network in the second case is $20$ times faster.
\section{Conclusion} \label{sec_conclusion}

We have presented a stability proof for the non-nominal circular limit cycles of the microgrid with a general type of voltage controller at each DER generator. The hurdle of the stability proof is the reference voltage, which is an input for the voltage controller and acts as a power source for the interconnected system. \textcolor{myblue}{Through a port-Hamiltonian model of the system with the definition of the time-varying phase $\hat\tau$, we formally remove the voltage sources from the energy balance equations and prove that the shifted passivity of the voltage controller, without considering the reference voltage or the swing equation, implies orbital stability.} The stability condition lends itself to simpler linear control design as a semidefinite program, whose effectiveness is demonstrated through simulation, showing superior performance of the control parameters tuned based on the proposed condition versus those tuned based on time-scale separation. Improvement in control performance includes (i) the guaranteed stability that is independent of the time constants of the distribution lines, and (ii) the uniform response in voltage and frequency such that the whole system operates on one time scale with smoother transients.




\bibliographystyle{plain}        
\bibliography{autosam}           



\nomenclature{$\matr M$}{Incidence matrix of the graph of the microgrid}
\nomenclature{$\matr M_1$}{Incidence matrix of the sub-graph of the microgrid that removes the ground node}
\nomenclature{$\vect u_{\texttt e, 1}$}{Edge input for interactions within network}
\nomenclature{$\vect y_{\texttt e, 1}$}{Edge output for interactions within network}
\nomenclature{$\matr G_{\texttt e, 1}$}{Input matrix for interactions within network}
\nomenclature{$\vect u_{\texttt e, 2}$}{Edge input for external interactions}
\nomenclature{$\vect y_{\texttt e, 2}$}{Edge output for external interactions}
\nomenclature{$\matr G_{\texttt e, 2}$}{Input matrix for external interactions}
\nomenclature{$\Upsilon_{\texttt e}(V_{\texttt e})$}{Load current}
\nomenclature{$\underline I$}{Vector of microgrid edge currents}
\nomenclature{$\underline V$}{Vector of microgrid edge voltages}
\nomenclature{$\underline U$}{Vector of microgrid node potentials}
\nomenclature{$\matr W$}{Network constraints}
\nomenclature{$L_c$}{DER coupling inductance}
\nomenclature{$R_c$}{DER coupling resistance}
\nomenclature{$C_f$}{DER $RLC$ filter capacitance}
\nomenclature{$L_f$}{DER $RLC$ filter inductance}
\nomenclature{$R_f$}{DER $RLC$ filter resistance}
\nomenclature{$\beta$}{First integrator state for DER control}
\nomenclature{$\xi$}{Second integrator state for DER control}
\nomenclature{$I_b$}{DER coupling line current}
\nomenclature{$V_o$}{DER output voltage}
\nomenclature{$I_f$}{DER terminal current}
\nomenclature{$V_t$}{DER terminal voltage}
\nomenclature{$m$}{DER inverter modulation index}
\nomenclature{$\omega$}{DER internal frequency}
\nomenclature{$\theta$}{DER internal swing angle}
\nomenclature{$R_{reg}$}{DER frequency regulation}
\nomenclature{$M$}{DER inertia constant}
\nomenclature{$V_\mathsf{nom}$}{Nominal DER voltage}
\nomenclature{$P_\mathsf{nom}$}{Nominal DER output power}
\nomenclature{$\omega_\mathsf{nom}$}{Nominal DER frequency}
\nomenclature{$D$}{Inverse of frequency regulation constant}
\nomenclature{$P_0$}{Output power at the projected zero frequency}
\nomenclature{$\matr K$}{DER control gain}
\nomenclature{$\epsilon$}{A scaling factor for the mechanical part of DER Hamiltonian}
\nomenclature{$\kappa$}{A scaling factor for the perturbation of the voltage controller dynamics}
\nomenclature{$\eqm\omega$}{Microgrid steady-state frequency}
\nomenclature{$\matr P$}{Perturbation for DER Hamiltonian}
\nomenclature{$\matr Q$}{Matrix for unperturbed DER Hamiltonian}
\nomenclature{$\tilde{\vect x}$}{Biased DER state vector}
\nomenclature{$\psi$}{$= -\kappa^{-1} V_\mathsf{nom}$}
\nomenclature{$\matr B$}{DER control input matrix}
\nomenclature{$K_{iv}$}{Parameter for second DER integrator}
\nomenclature{$\matr E_1$}{A matrix that remove null space}
\nomenclature{$\matr N$}{Matrix that combines real and imaginary parts}
\nomenclature{$\matr A_{\texttt e}$}{Electrical part of DER dissipation}
\nomenclature{$\vect b_{\texttt e}$}{Cross terms of DER dissipation}
\nomenclature{$\vect s$}{Co-energy state vector of the microgrid}
\nomenclature{$\nu$}{$= (\omega - \eqm\omega)/2$}
\nomenclature{$\vartheta$}{DER internal swing angle in $\frac{\omega}{2}$--reference frame}
\nomenclature{$\texttt g$}{Number of DER generators in the microgrid}
\nomenclature{$\texttt T$}{Number of $R$--$L$ lines in the microgrid}
\nomenclature{$\ell$}{Number of loads in the microgrid}
\nomenclature{$\texttt n$}{Total number of edges in the graph of the microgrid}
\nomenclature{$\eqm{\vect x}(\tau)$}{Limit cycle solution with phase angle $\tau$}
\nomenclature{$\texttt e$}{Index for edge number}
\nomenclature{$\matr K_1$}{A reformation of $\matr K$}


\end{document}